\documentclass{article}

\usepackage{arxiv}

\usepackage[utf8]{inputenc} %
\usepackage[T1]{fontenc}    %
\usepackage{hyperref}       %
\usepackage{url}            %
\usepackage{booktabs}       %
\usepackage{amsfonts}       %
\usepackage{nicefrac}       %
\usepackage{microtype}      %
\usepackage{lipsum}		%
\usepackage{graphicx}
\usepackage{natbib}
\usepackage{doi}
\usepackage[ruled,vlined]{algorithm2e}
\usepackage{enumitem}

\usepackage{amsmath, amssymb, amsfonts, amsthm}
\usepackage{bm}

\usepackage{xcolor}

\title{Bounding the Null Space:\linebreak Interval-Based Uncertainty Quantification\linebreak for Non-Identifiable Groundwater Models}

\author{ \href{https://orcid.org/0000-0003-3508-6214}{\includegraphics[scale=0.06]{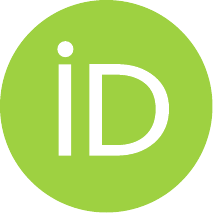}\hspace{1mm}Maximilian Ramgraber}\\
	Department of Geoscience \& Engineering\\
	Delft University of Technology\\
	Delft, 2628 CN \\
	\texttt{m.ramgraber@tudelft.nl} \\
	\AND
	\href{https://orcid.org/0000-0002-7018-7099}{\includegraphics[scale=0.06]{orcid.pdf}\hspace{1mm}Ksenia Bestuzheva} \\
	Zuse Institute Berlin \\
	Berlin, 14195 \\
	\texttt{bestuzheva@zib.de} \\
}

\renewcommand{\shorttitle}{Bounding the Null Space}
\newcommand{\dx}{\operatorname{dx}}
\newcommand{\dhx}{\operatorname{dhx}}

\newcommand{\dht}{\operatorname{dht}}

\definecolor{blue_custom}{HTML}{1988B8}
\definecolor{orange_custom}{HTML}{FF5000}
\definecolor{green_custom}{HTML}{01A109}
\newcommand{\cb}[1]{\textcolor{blue_custom}{#1}}
\newcommand{\co}[1]{\textcolor{orange_custom}{#1}}

\hypersetup{
    hidelinks,
    pdfsubject={Computational physics; groundwater uncertainty quantification}
}

\hypersetup{
pdftitle={Bounding the Null Space: Interval-Based Uncertainty Quantification for Non-Identifiable Groundwater Models},
pdfauthor={Maximilian~Ramgraber, Ksenia~Bestuzheva},
pdfkeywords={Hydrogeology, Uncertainty Quantification, Linear Programming, Null Space, Nonidentifiability, Constraint Propagation},
}

\begin{document}
\maketitle

\begin{abstract}
Groundwater models are routinely non-identifiable: sparse subsurface observations leave many combinations of parameters, states, and boundary conditions equally consistent with the available data. Existing uncertainty quantification (UQ) methods address this by exploring a finite set of model realizations, but incomplete exploration can systematically underestimate the true range of admissible solutions. We propose a fundamentally different approach based on Optimization-based Bound Tightening (OBBT), which represents uncertainty directly as intervals and tightens them by extremizing variables over a constraint system encoding physical laws and observations. This yields guaranteed outer bounds on all uncertain variables without sampling, side-stepping the exploration problem entirely. To apply OBBT to groundwater flow, we discretize Darcy's law using a finite-volume scheme and handle the resulting bilinear terms through McCormick relaxations. We show that these relaxations can break the sign coupling between fluxes and head gradients, permitting non-physical rotational flow and failing to provide sufficient information for effective bound tightening. We identify flow sign prescription and irrotationality constraints as effective remedies and characterize their respective strengths and limitations. We demonstrate the framework on three numerical examples — a 1D steady-state model, a 2D steady-state model across four experimental configurations, and a 2D transient model on a hexagonal grid — and discuss computational performance, scalability, and directions for future research. OBBT offers a conservative, deterministic, and physically grounded alternative to ensemble-based UQ, with natural connections to null space theory and data assimilation.
\end{abstract}


\section{Introduction}\label{sec:introduction}

Groundwater modeling must contend with uncertain hydraulic states, aquifer properties, and boundary conditions. One persistent source of this uncertainty is \textbf{equifinality} \citep[also called \textit{non-uniqueness}, \textit{non-identifiability}, or \textit{indeterminacy}; e.g.,][]{beven2001equifinality,beven2006manifesto}. Because subsurface observations are sparse, the available data often do not support a unique description of the system. Different combinations of parameters and boundary conditions can explain the same observations. Uncertainty therefore remains even after calibration.

Exact \textbf{uncertainty quantification} (UQ) is computationally feasible only in simple special cases. In practice, approximate methods must risk either understating or overstating uncertainty. In groundwater applications, the more serious error is usually understatement \citep{doherty2020decision,middlemis2017groundwater,merz2012australian}, as risks often arise from rare extremes such as droughts or floods. Conservative UQ is therefore desirable for safe and reliable groundwater management.

Most established UQ methods represent uncertainty by exploring a finite set of possible system realizations. Some examine the neighborhood of a calibrated optimum, as in regularized inversion and sensitivity-based workflows \citep[e.g., ][]{tonkin2009calibration}. Others rely on ensembles or samples of alternative realizations, as in Monte Carlo methods, ensemble data assimilation, or Bayesian posterior sampling \citep[e.g., ][]{hendricks2008real,ramgraber2021non,laloy2013efficient}. These approaches differ in formulation, but they share the same premise: uncertainty is inferred from a limited exploration of a much larger feasible space. This paradigm is both natural and widely used, but it also ties the quality of UQ to the extent and fidelity of that exploration. 

This creates a structural dilemma. The more complex the system, the more unknown parameters, the more possibilities we must explore. Yet, the more complex the system, the more expensive the simulation, the fewer possibilities we can explore. In other words: as the need for exploration grows, our ability for exploration shrinks. When uncertainty is represented through incomplete exploration, under-exploration can become underestimation of uncertainty.

The problem is therefore one of representation. For conservative UQ, we may want to capture the extent of the feasible region itself, for example through outer bounds on uncertain variables. Yet any empirical bounds derived from a finite sample can only underestimate the true extent of that region. This limitation motivates a different representation of uncertainty.

One alternative is to represent uncertainty directly by intervals. Instead of enumerating distinct realizations, we can work with admissible value ranges implied by prior bounds, physical laws, and observations. This perspective is especially attractive under non-identifiability. Initial bounds may be broad, and physics and data can then reduce them conservatively. If the system is non-identifiable, the marginal bounds tighten around the set of equivalent solutions, that is, the \textbf{null manifold} (Section~\ref{subsec:null_space}).

The technical realization of this idea is \textbf{Optimization-based bound tightening} \citep[OBBT: e.g.,][]{gleixner2017three}. While OBBT is established in mathematical programming, its application to physically-constrained groundwater UQ has not previously been explored. In OBBT, we formulate (in)equality constraints from prior bounds, observations, and governing physics. We then extremize each variable over the resulting \textbf{mathematical program} to compute conservative lower and upper bounds on its admissible values.

For hydrogeological UQ, applying OBBT to groundwater models requires rewriting the governing partial differential equations as a system of linear constraints. Constraint-based treatments of PDEs have received increasing attention in recent years \citep[e.g.,][]{gnegel2021solution,leyffer2025mccormick}, but nonlinear PDEs -- such as Darcy flow under parameter uncertainty -- remain difficult.  
A possible solution is relaxing the governing nonlinear partial differential equations via a system of linear constraints \citep{leyffer2025mccormick,mccormick1976computability}.

In the following, we formulate discretized groundwater flow as a constraint system and apply optimization-based bound tightening to compute admissible ranges for hydraulic heads, fluxes, recharge, and parameters. We examine how McCormick-based linearization affects groundwater flow physics, how these effects can be mitigated, and demonstrate the resulting algorithm over a range of numerical examples. 

\section{Background}\label{sec:background}

\subsection{Constraint systems}\label{subsec:constraint_systems}

In the introduction, we advocated for a different representation of uncertainty. Rather than enumerating possibilities, we recommended marginal variable intervals that represent admissible value ranges for an uncertain variable $x$ given initial bounds, physics, and data.

\begin{figure}
  \centering
  \includegraphics[width=\textwidth]{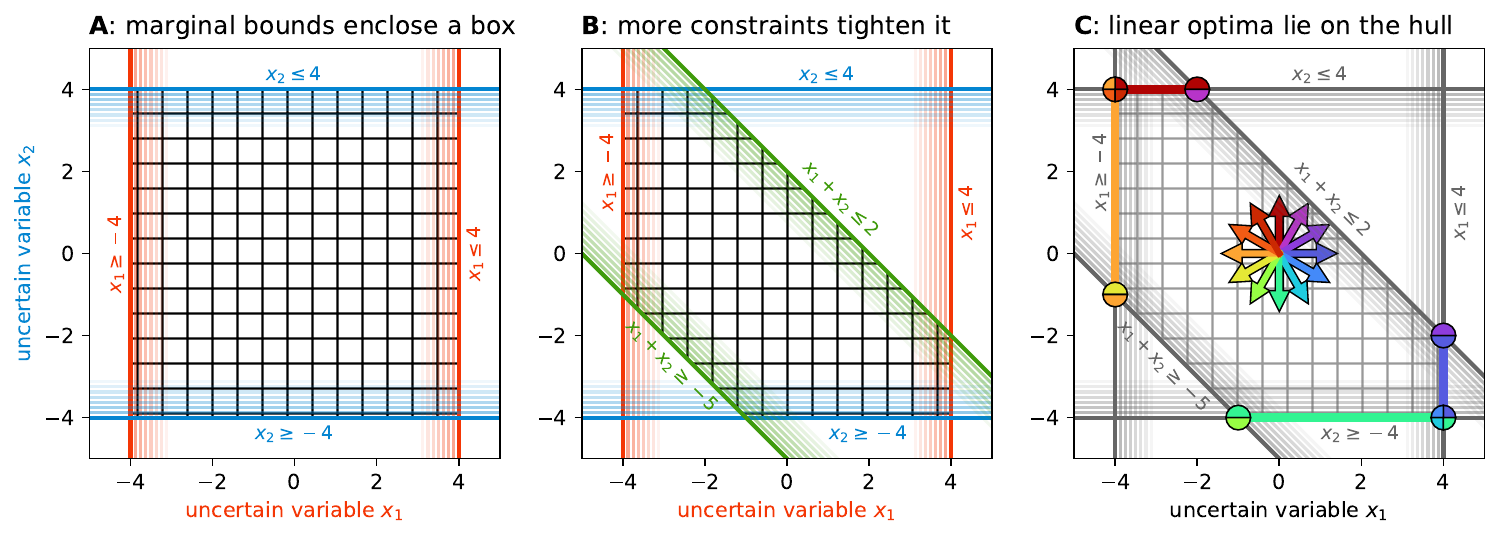}
  \caption{(A) Marginal bounds define a (hyper)box in variable space. The hatched area denotes the feasible region. (B) Additional constraints tighten the feasible region. (C) Linear optima of a linear constraint system lie on the hull of its feasible region. Color denotes different linear objectives (arrows) and their optima (edges and vertex wedges).}
  \label{fig:feasible_region}
\end{figure}

Let $x_1,x_2 \in \mathbb{R}$ be uncertain variables with initial bounds $x_{i}\in[\underline{x_{i}},\overline{x_{i}}]$, such that $\underline{x_{i}} \leq x_{i} \leq \overline{x_{i}}$ for $i=1,2$. Each linear inequality $\underline{x_{i}} \leq x_{i}$ or $x_{i} \leq \overline{x_{i}}$ of these marginal variable bounds defines an axis-aligned half-space in a (here: two-dimensional) variable space. Together, these inequality constraints enclose a (hyper)box of points $\boldsymbol{x}:=(x_1,x_2)\in\mathbb{R}^2$ that fulfill all prescribed inequalities, the so-called \textbf{feasible region}. Prescribing further inequality constraints can tighten the feasible region. An example of this is shown in Figure~\ref{fig:feasible_region}A and B. 

If all constraints are linear, the enclosed feasible region is always a convex polytope. This has elegant properties for optimization. Recall that a linear optimization objective corresponds to descending in a global direction in variable space. Since the feasible space is convex, a global optimum will always lie at an extremum of the polytope, usually one of its vertices. If more than one vertex carries a global optimum, all points on the edge or face between these vertices will also be a global optimum (Figure~\ref{fig:feasible_region}C). This makes the resulting optimization problem highly efficient: the optimization problem has a nice, well-defined structure, and is convex. Optimization strategies which exploit these properties are known as \textbf{linear programming} (LP).

Solvers for linear programs are usually more efficient and reliable than solvers for many more general convex optimization problems due to the additional structure imposed by the constraints. Simplex methods, for example, exploit the fact that in fully bounded problems there is always an optimal vertex. While their worst-case computational complexity is exponential, 
this pathological behavior is fragile, and their expected computational cost is polynomial for problem data with small numerical variability \citep{spielman2004smoothed}. 

\subsection{Non-identifiability}\label{subsec:null_space}

In the previous section, we have discussed constraint systems defined by (linear) inequalities. We saw how additional constraints can tighten a feasible region. In most physical systems, however, the governing equations are defined as equalities, not inequalities. In constraint systems, equalities can be represented through two opposing inequalities. For instance, the equality $x_1 + x_2 = 1$ can be represented through two inequalities $x_1 + x_2 \leq 1$ and $x_1 + x_2 \geq 1$. The feasible region then reduces to the only part of the variable space that satisfies both constraints: the line $x_1 + x_2 = 1$. Visually, this would correspond to closing the gap between the two diagonal green inequalities in Figure~\ref{fig:feasible_region}B.

In general, if the resulting feasible region is not collapsing to a point, then there are multiple solutions that satisfy the system of constraints. The consequence is non-identifiability. In the following, we connect this constraint view to the concepts of null spaces and null manifolds, which offer a different perspective on non-identifiability.

\subsubsection{Null spaces}

 Consider a linear system of equations of the form $\boldsymbol{A}\boldsymbol{x}=\boldsymbol{y}$, where 
\begin{itemize}
    \item $\boldsymbol{A}$ is a linear operator (an $O$-by-$D$ matrix),
    \item $\boldsymbol{x}$ is a vector of input variables (a $D$-by-$1$ vector), and 
    \item $\boldsymbol{y}$ is a vector of output variables (an $O$-by-$1$ vector).
\end{itemize}
If $D>O$, the system has nonzero degrees of freedom and the inverse problem is non-identifiable. In linear algebra, this indeterminacy can be revealed elegantly through a singular value decomposition (SVD) of $\boldsymbol{A}$. Assuming $\operatorname{rank}(\boldsymbol{A})=O$, we have:
\begin{equation}
    \begin{aligned}
    \boldsymbol{A} = \boldsymbol{U}\boldsymbol{S}\boldsymbol{V}^\intercal = \overbrace{\left[
    \begin{matrix}
        \cb{\boldsymbol{u}_1} & \cb{\boldsymbol{u}_2} & \cb{\cdots} & \cb{\boldsymbol{u}_{O}}
    \end{matrix}
    \right]}^{\boldsymbol{U}:\; O \times O\text{ matrix}}    &\overbrace{\left[
    \begin{matrix}
        \cb{s_1} & \cb{0} & \cb{0} & \cb{\cdots} & \cb{0} & \co{0} & \co{\cdots} & \co{0} \\
        \cb{0} & \cb{s_2} & \cb{0} & \cb{\cdots} & \cb{0} & \co{0} & \co{\cdots} & \co{0}\\
        \cb{0} & \cb{0} & \cb{s_3} & \cb{\cdots} & \cb{0} & \co{0} & \co{\cdots} & \co{0}\\
        \cb{\vdots} & \cb{\vdots} & \cb{\vdots} & \cb{\ddots} & \cb{\vdots} & \co{\vdots} & \co{\vdots} & \co{\vdots}\\
        \cb{0} & \cb{0} & \cb{0} & \cb{\cdots} & \cb{s_{O}} & \co{0} & \co{\cdots} & \co{0}\\
    \end{matrix}
    \right]}^{\boldsymbol{S}:\;O \times D\text{ matrix}}
    \overbrace{\left[
    \begin{matrix}
        \cb{\boldsymbol{v}_1}^\intercal \\
        \cb{\vdots}\\
        \cb{\boldsymbol{v}_{O}}^\intercal \\
        \co{\boldsymbol{v}_{O+1}}^\intercal \\
        \co{\vdots}\\
        \co{\boldsymbol{v}_{D}}^\intercal 
    \end{matrix}\right]}^{\boldsymbol{V}^\intercal:\;D \times D\text{ matrix}},\\
    &\qquad\cb{\text{active space}}\qquad\qquad \co{\text{null space}}
    \end{aligned}
    \label{eq:singular_value_decomposition}
\end{equation}
where $\boldsymbol{u}_{i}$ is the $i$-th left singular vector (a $O \times 1$ vector), $\boldsymbol{v}_{j}$ is the $j$-th right singular vector (a $D \times 1$ vector), and $s_k$ is the $k$-th singular value (a scalar). We note three things:
\begin{enumerate}
    \item The left singular vectors $\boldsymbol{u}_{1}$ to $\boldsymbol{u}_{O}$ define a new orthonormal basis in the output space, in which $\boldsymbol{y}$ lives.
    \item The right singular vectors $\boldsymbol{v}_{1}$ to $\boldsymbol{v}_{D}$ define a new orthonormal basis in the input space, in which $\boldsymbol{x}$ lives.
    \item The singular value matrix $\boldsymbol{S}$ is diagonal with $s_1 \geq \cdots \geq s_O \geq 0$.
\end{enumerate}
If we take a closer look at the matrix product $\boldsymbol{S}\boldsymbol{V}^\intercal$ in Equation~\ref{eq:singular_value_decomposition}, we can see that the last $D-O$ right singular vectors $\boldsymbol{v}_{O+1},\dots,\boldsymbol{v}_{D}$ are multiplied with the all-zero columns of $\boldsymbol{S}$. Since a multiplication by zero eliminates the contribution of these vectors to the final result, the subspace spanned up by these last right singular vectors is referred to as the \textbf{null~space} \citep[e.g., ][]{doherty2011approaches,doherty2020decision}. In Equation~\ref{eq:singular_value_decomposition}, we have marked the entries associated with the null space in \co{orange}, and entries associated with its complement (the \textbf{active space}) in \cb{blue} .

\begin{figure}
  \centering
  \includegraphics[width=\textwidth]{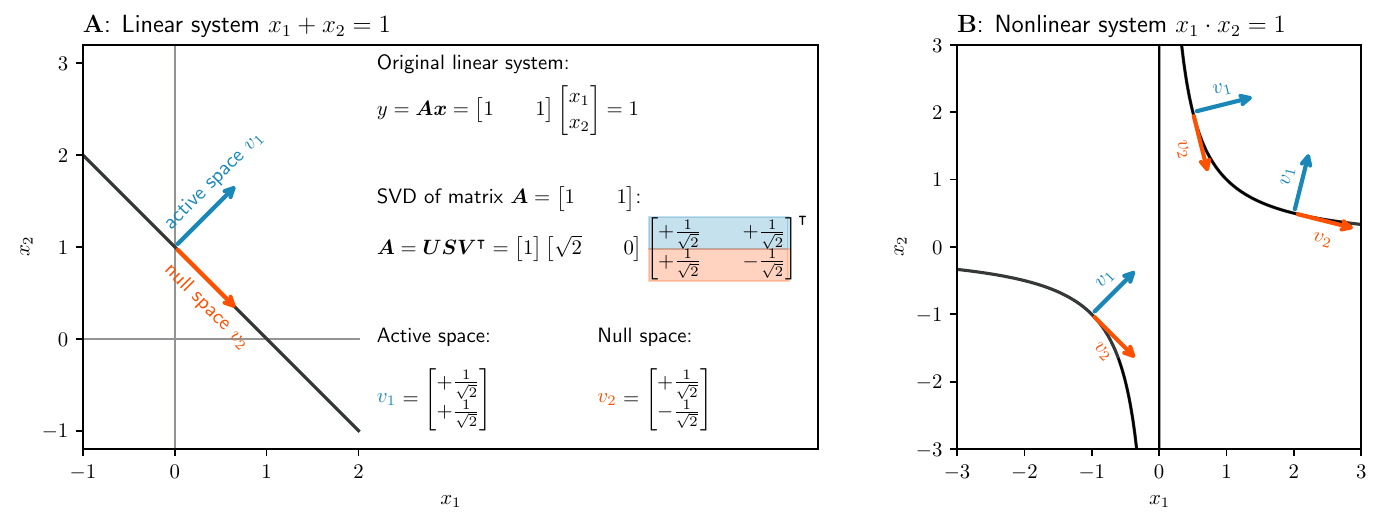}
  \caption{(A) The linear system $x_1 + x_2 = 1$ is non-identifiable and thus has an active space and a null space. An SVD reveals the vectors that span both spaces. The null space is aligned with the set of equivalent solutions (black diagonal line), the active space lies orthogonal to it. Moving along the null space dimension $v_2$ increases $x_1$ by the same amount that it decreases $x_2$, or vice versa, preserving the value of the sum. (B) In a nonlinear system, equivalent solutions lie along a curved manifold. While we can still derive \textit{local} null space vectors as tangents of this manifold, these vectors do not extrapolate globally.}
  \label{fig:null_space}
\end{figure}

In practical terms, the existence of a null space implies that there exist combinations of input variables $\boldsymbol{x}$ that cancel out and thus do not affect the value of the output variables $\boldsymbol{y}$. In mathematical terms, the SVD splits the input vector $\boldsymbol{x}$ into active and null space components, of which only the active space components affect the result:
\begin{equation}
    \begin{aligned}
        \boldsymbol{y}
        &= \boldsymbol{A}\boldsymbol{x} \\
        &= \boldsymbol{U}\boldsymbol{S}\boldsymbol{V}^\intercal\boldsymbol{x} \\ 
        &= 
        \overbrace{\boldsymbol{U}}^{O \times O}
        \left[
        \begin{matrix}
            \overbrace{\cb{\boldsymbol{S}_{a}}}^{O \times O}
            &
            \overbrace{\co{\boldsymbol{0}}}^{O \times (D-O)}
        \end{matrix}
        \right]
        \left[
        \begin{matrix}
            \overbrace{\cb{\boldsymbol{V}_{a}^{\intercal}}}^{O \times D} \\
            \overbrace{\co{\boldsymbol{V}_{n}^{\intercal}}}^{(D-O) \times D}
        \end{matrix}
        \right]
        \overbrace{\boldsymbol{x}}^{D \times 1} \\
        &=
        \boldsymbol{U}
        \left(
            \underbrace{
            \overbrace{
            \cb{\boldsymbol{S}_{a}\boldsymbol{V}_{a}^{\intercal}\boldsymbol{x}}
            }^{O \times 1}
            }_{\text{\cb{active space} contribution}}
            +
            \underbrace{
            \overbrace{
            \co{\boldsymbol{0}\boldsymbol{V}_{n}^{\intercal}\boldsymbol{x}}
            }^{O \times 1}
            }_{\text{\co{null space} contribution}}
        \right) \\
        &=
        \boldsymbol{U}
        \underbrace{
        \overbrace{
        \cb{\boldsymbol{S}_{a}\boldsymbol{V}_{a}^{\intercal}\boldsymbol{x}}
        }^{O \times 1}
        }_{\text{\cb{active space} contribution}}.
    \end{aligned}
    \label{eq:null_space_split}
\end{equation}
where $\cb{\boldsymbol{S}_{a}}=\operatorname{diag}(s_1,\ldots,s_O)$,
$\cb{\boldsymbol{V}_{a}}=[\boldsymbol{v}_1,\ldots,\boldsymbol{v}_O]$, and
$\co{\boldsymbol{V}_{n}}=[\boldsymbol{v}_{O+1},\ldots,\boldsymbol{v}_D]$.
Equation~\ref{eq:null_space_split} shows that only the coordinates of
$\boldsymbol{x}$ along the active-space basis $\cb{\boldsymbol{V}_{a}}$
contribute to $\boldsymbol{y}$. The coordinates along the null space basis
$\co{\boldsymbol{V}_{n}}$ are multiplied by the zero block in
$\boldsymbol{S}$ and are therefore eliminated. Consequently, any perturbation
$\Delta\boldsymbol{x}$ that lies entirely in the null space, i.e.,
$\Delta\boldsymbol{x}=\co{\boldsymbol{V}_{n}}\boldsymbol{\beta}$ for some
coefficient vector $\boldsymbol{\beta}$, leaves the output unchanged:
$\boldsymbol{A}(\boldsymbol{x}+\Delta\boldsymbol{x})=\boldsymbol{A}\boldsymbol{x}$.
This makes the inverse problem, estimating $\boldsymbol{x}$ from
$\boldsymbol{y}$, non-identifiable. A minimal example of this principle is illustrated in Figure~\ref{fig:null_space}A.

We can juxtapose two complementary perspectives on non-identifiability for a given system of (in)equalities:
\begin{enumerate}
    \item In \textbf{linear programming}, the system of constraints defines a feasible region in variable space. Non-identifiability arises when this region has nonzero dimension, such that multiple points satisfy all constraints.
    \item In \textbf{inverse modelling}, non-identifiability arises when multiple parameter vectors $\boldsymbol{x}$ produce the same observations $\boldsymbol{y}$. In the linear case, this corresponds to the existence of a nontrivial null space.
\end{enumerate}
Both approaches describe the same principle from different sides: The feasible region in constraint space coincides with the set of solutions to $\boldsymbol{A}\boldsymbol{x}=\boldsymbol{y}$, while the null space characterizes directions along which this solution set can be traversed without affecting $\boldsymbol{y}$. Thus, the dimension of the feasible region is directly linked to the dimension of the null space. In this sense, non-identifiability is geometric in the linear programming view and algebraic in the inverse modelling view, but reflects the same underlying non-uniqueness of solutions.

\subsubsection{Null manifolds}

Indeterminacy is not limited to linear models. As such, a similar concept to null spaces also generalizes to nonlinear systems of equations. In such nonlinear settings, the null space will no longer be a Euclidean subspace of the original input space. Instead, the feasible region will lie on some curved manifold (Figure~\ref{fig:null_space}B). Terminology for this more general concept varies between disciplines. It is sometimes referred to as the "kernel", the "solution set", the "pre-image of a model". Reflecting the nomenclature we adopted in the introduction, we will refer to the nonlinear generalization of a null space as a \textbf{null manifold}. 

Note that it is still possible to derive approximate local linear null space vectors by first linearizing a nonlinear system, then identifying the null space of that linearized system. Such approaches are known as \textbf{null space Monte Carlo} and have found successful application in hydrogeology \citep[e.g., ][]{doherty2011approaches}. For the purpose of UQ, however, we would prefer globally valid enclosures of the null manifold, which we will explore in the following.

\subsection{Optimization-based bound tightening}\label{subsec:OBBT}

\begin{figure}
  \centering
  \includegraphics[width=\textwidth]{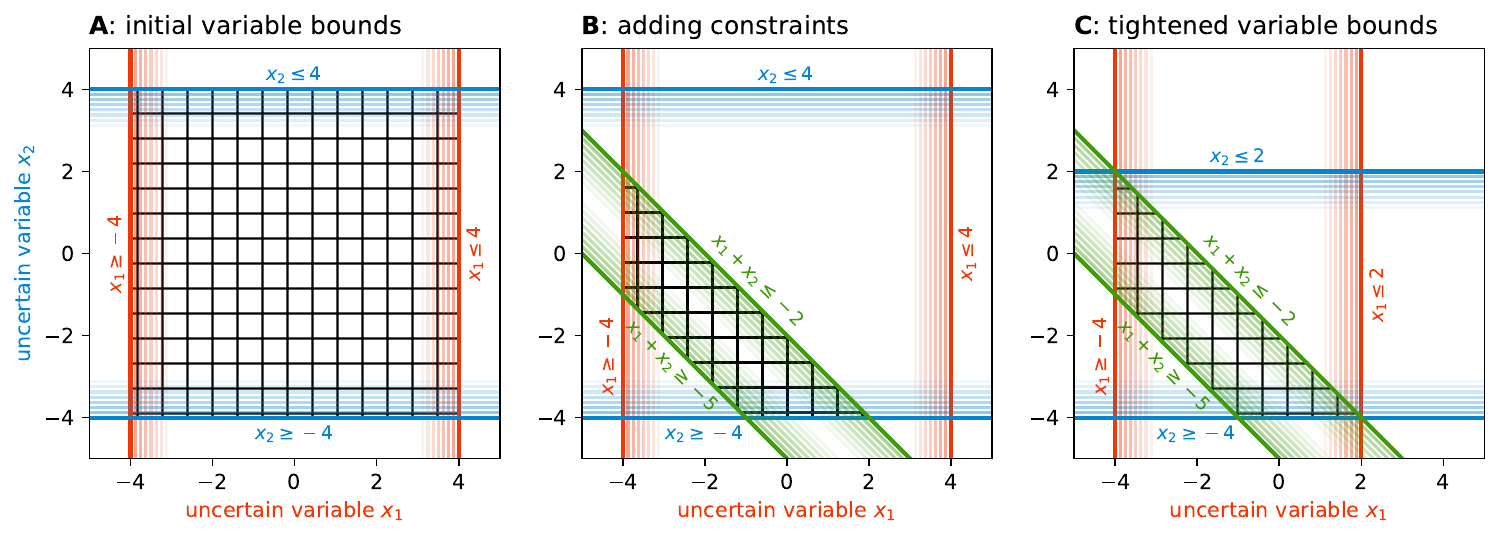}
  \caption{Schematic illustration of OBBT. (A) Each marginal variable has initial bounds. Together, the feasible region defines a box. (B) Additional constraints narrow the feasible region. Some constraints (\textcolor{orange_custom}{$x_1 \leq 4$} and \textcolor{blue_custom}{$x_2 \leq 4$}) may become redundant. (C) Solving LPs to maximize and minimize each marginal variable identifies and tightens redundant marginal variable bounds (\textcolor{orange_custom}{$\overline{x_1}: 4 \rightarrow 2$} and \textcolor{blue_custom}{$\overline{x_2}: 4 \rightarrow 2$}).}
  \label{fig:OBBT}
\end{figure}

In an interval- or constraint-based representation of uncertainty, inference
corresponds to narrowing the admissible value intervals of uncertain variables.
This can be achieved through optimization-based bound tightening (OBBT). The
key idea is as follows: Initial variable bounds define a (hyper-)rectangular
feasible region in variable space. These bounds encode the initial admissible
ranges, analogous to prior uncertainty estimates (Figure~\ref{fig:OBBT}A).
Additional constraints, such as physics, data, and prior assumptions, further
restrict the feasible region. As a result, some initial bounds may become
redundant (Figure~\ref{fig:OBBT}B). By extremizing each variable over the
feasible region, we can identify and remove infeasible portions
of the initial bounds.

To tighten the bounds of a variable of interest in a linear constraint system,
we solve two LPs per variable: one to minimize the variable and one to maximize
it, subject to all constraints. For variable $x_i$, these problems are
\begin{equation}
    \underline{x}_i^{\,\mathrm{tight}}
    =
    \min x_i
    \quad \text{subject to } \boldsymbol{x}\in\mathcal{F},
    \qquad
    \overline{x}_i^{\,\mathrm{tight}}
    =
    \max x_i
    \quad \text{subject to } \boldsymbol{x}\in\mathcal{F},
    \label{eq:obbt_bounds}
\end{equation}
where $\mathcal{F}$ denotes the feasible region defined by the current variable
bounds and all additional constraints. If the minimization yields a value higher
than the current lower bound, the lower bound can be raised:
$\underline{x}_i\leftarrow
\max(\underline{x}_i,\underline{x}_i^{\,\mathrm{tight}})$.
Likewise, if the maximization yields a value lower than the current upper bound,
the upper bound can be lowered:
$\overline{x}_i\leftarrow
\min(\overline{x}_i,\overline{x}_i^{\,\mathrm{tight}})$.

This procedure removes value ranges that cannot occur in any feasible solution
and therefore tightens the marginal bounds of each variable. In the context of
non-identifiability, these tightened bounds describe the marginal extent of the
set of admissible solutions, or null manifold, along each coordinate direction
(Figure~\ref{fig:OBBT}C). Thus, OBBT provides guaranteed lower and upper bounds
on each variable of interest, even when the inverse problem remains
non-identifiable.

\section{Darcy flow}\label{sec:Darcy_flow}

With this, we have a general framework for state and parameter inference under uncertainty. In this chapter, we introduce the discretized Darcy flow that will provide the physical constraints for the numerical experiments explored in Section~\ref{sec:examples}.

\subsection{Discretized groundwater flow}\label{subsubsec:groundwater_flow}

In this study, we focus on groundwater flow governed by Darcy's law. In continuous form, the mass balance can be written as
\begin{equation}
\begin{aligned}
\nabla \cdot \mathbf{q}(\mathbf{s}) - R(\mathbf{s}) &= 0
&& \quad \text{steady-state},\\
\underbrace{S(\mathbf{s})\,\frac{\partial h(\mathbf{s},t)}{\partial t}}_{\text{storage change}}
+
\underbrace{\nabla \cdot \mathbf{q}(\mathbf{s},t)}_{\text{net spatial outflow}}
-
\underbrace{R(\mathbf{s},t)}_{\text{net source term}}
&= 0
&& \quad \text{transient}.
\end{aligned}
\label{eq:pde_mass_balance}
\end{equation}
where $\boldsymbol{s}$~[m] is a spatial location, $t$~[s] is a time, $h$~[m] is hydraulic head, $\mathbf{q}$~[m²/s] is the specific discharge integrated over aquifer thickness (hereafter referred to as ``flux''), $R$~[m/s] denotes recharge (or, more generally, net sources/sinks per unit area), and $S$~[-] denotes the specific yield. $\nabla$ is the spatial derivative operator $\nabla = [\frac{\partial}{\partial s_1}, \dots, \frac{\partial}{\partial s_k}]$ in $k \leq 3$ spatial dimensions, and $\nabla \cdot \mathbf{q}(\mathbf{s},t) = \frac{\partial q_{s_1}}{\partial s_1} + \dots + \frac{\partial q_{s_k}}{\partial s_k}$~[m/s] consequently is the divergence of the flow.
We define the specific discharge with Darcy's law,
\begin{equation}
\mathbf{q}(\mathbf{s},t) = -T(\mathbf{s},t)\,\nabla h(\mathbf{s},t),
\label{eq:pde_darcy}
\end{equation}
where $T$~[m²/s] is transmissivity. For fixed $T$, Equation~\ref{eq:pde_darcy} is linear in $h$. However, if both $T$ and $\nabla h$ are uncertain, the resulting flux relations become bilinear in both variables. To obtain algebraic relations suitable for OBBT, we must first discretize the system of equations \citep{gnegel2021solution}.

To this end, we adopt a finite-volume scheme for Equation~\ref{eq:pde_mass_balance}. Let cell $i$ have area $A_i$~[m²] and a set of neighboring cells $\operatorname{neighbours}(i)$. Integrating \eqref{eq:pde_mass_balance} over cell $i$ and applying the divergence theorem yields the discrete mass balance
\begin{equation}
\begin{aligned}
\sum_{j \in \operatorname{neighbours}(i)} q_{j \rightarrow i}^{t} - A_i R_{i}^{t} &= 0 && \quad\text{steady-state},\\
\sum_{j \in \operatorname{neighbours}(i)} q_{j \rightarrow i}^{t} - A_i R_{i}^{t} &= q_{i}^{t-1 \rightarrow t} && \quad\text{transient},
\end{aligned}
\label{eq:mass_balance}
\end{equation}
where $q_{j \rightarrow i}^{t}$~[m³/s] denotes the volumetric flux entering cell $i$ across the interface shared with neighbor $j$ at timestep $t$. The transient formulation additionally includes a storage term $q_{i}^{t-1 \rightarrow t}$~[m³/s], which can be conceptually parsed as a flow across time in a spatio-temporal grid. We parameterize the interface fluxes using a two-point flux approximation. Specifically, we define
\begin{equation}
\begin{aligned}
q_{j \rightarrow i}^{t} &= T_{j,i}\,\dhx_{j \rightarrow i}^{t}\,w_{j,i} && \quad\text{spatial flux},\\
q_{i}^{t-1 \rightarrow t} &= S_{i}\,\dht_{i}^{t}\,A_{i} && \quad\text{storage term},
\end{aligned}
\label{eq:Darcys_law}
\end{equation}
where $T_{j,i}$ is an effective transmissivity associated with the interface $(j,i)$, $w_{j,i}$ denotes the flow-active interface measure (e.g., face length in 2D), and the discrete gradients are
\begin{equation}
\begin{aligned}
\dhx_{j \rightarrow i}^{t} &= \frac{h_{j} - h_{i}}{\Delta x_{j,i}} && \quad\text{head gradient in space},\\
\dht_{i}^{t} &= \frac{h_{i}^{(t)} - h_{i}^{(t-1)}}{\Delta t} && \quad\text{head gradient in time}.
\end{aligned}
\label{eq:head_gradient}
\end{equation}

Equations~\ref{eq:mass_balance} to \ref{eq:head_gradient} highlight three properties that are central to the bound-tightening strategy discussed in Section~\ref{sec:background}:
\begin{enumerate}
    \item \textbf{Linear mass balance.} For fixed grid geometry (i.e., fixed $A_i$, $w_{j,i}$, and $\dx_{j,i}$), the mass balance equations~\ref{eq:mass_balance} are linear in the fluxes and source terms.
    \item \textbf{Nonlinear Darcy equalities.} The constitutive flow definitions (Equation~\ref{eq:Darcys_law}) are bilinear when both $T_{j,i}$ and $\dhx_{j\to i}$ are unknown. In transient settings, the storage term is likewise bilinear when both $S_i$ and $\dht_i$ are unknown. The width $w_{j,i}$ and surface area $A_{i}$ are defined by grid geometry and are thus generally constants.
    \item \textbf{Linearity for known properties.} If either the head field ($h_i^t$), the gradient field ($\dhx_{j \rightarrow i}^{t}$ and $\dht_{i}^{t}$), or the material properties ($T_{j,i}$ and $S_i$) are known, the corresponding relations become linear.
\end{enumerate}
For the purposes of hydrogeological inference, subsurface properties
($T$ and $S$), forcings ($R$), and states ($h$) are all generally uncertain.
This results in bilinear terms that make the constraint system nonlinear and
nonconvex. If treated exactly, the resulting bound-tightening subproblems are
therefore nonconvex nonlinear optimization problems. While nonlinear programming
solvers exist and can be used for OBBT, there are strong computational and
methodological reasons to prefer linear or convex formulations
wherever possible.

Local nonlinear programming solvers such as IPOPT
\citep[][]{wachter2006implementation} generally only return local optima. This is
problematic for OBBT, because guaranteed bounds require the minimization and
maximization problems to be solved globally. Global optimization solvers such as
SCIP \citep{bolusani2024scip} can provide such guarantees, but typically at
substantially higher computational cost. Since discretized groundwater systems
often have thousands of unknown variables, efficient solution schemes can make
the difference between practical usefulness and computational infeasibility.
This motivates the use of linear relaxations that yield conservative bounds
while allowing the subproblems to be solved with LP solvers.

\subsection{McCormick relaxations}\label{subsec:McCormick}

Fortunately, it is possible to linearize bilinear terms and thereby leverage the computational efficiency of linear programs. This can be achieved with the help of \textbf{McCormick relaxations} \citep{mccormick1976computability}. Let us consider a bilinear product of the form
\begin{equation}
    w = x y,
\end{equation}
where $x \in [\underline{x},\overline{x}]$ and $y \in [\underline{y},\overline{y}]$ are two variables with known finite lower and upper bounds. Then, a McCormick relaxation replaces this bilinear equality constraint with four linear inequalities:
\begin{equation}
\begin{aligned}
w &\ge \underline{x}\,y + \underline{y}\,x - \underline{x}\,\underline{y} && 1^{\text{st}} \text{ underestimator} \\
w &\ge \overline{x}\,y + \overline{y}\,x - \overline{x}\,\overline{y} && 2^{\text{nd}} \text{ underestimator}\\
w &\le \overline{x}\,y + \underline{y}\,x - \overline{x}\,\underline{y} && 1^{\text{st}} \text{ overestimator} \\
w &\le \underline{x}\,y + \overline{y}\,x - \underline{x}\,\overline{y} && 2^{\text{nd}} \text{ overestimator,}
\end{aligned}    
\label{eq:McCormick}
\end{equation}
where the first two inequalities are \textbf{underestimators} for $w$, and the last two inequalities are \textbf{overestimators} for $w$. Together, these under- and overestimators constitute an (optimally-tight) linear relaxation of the original bilinear constraint: a \textbf{McCormick envelope}. This means that the McCormick relaxation forms a convex hull around the original nonlinear feasible region. This makes the feasible region more permissible than the original bilinear constraint, but this is important for the purpose of reliable UQ: if we introduce approximations, we want to ensure that they will not result in an underestimation of uncertainty. A relaxation ensures that this will not happen.

\begin{figure}
  \centering
  \includegraphics[width=\textwidth]{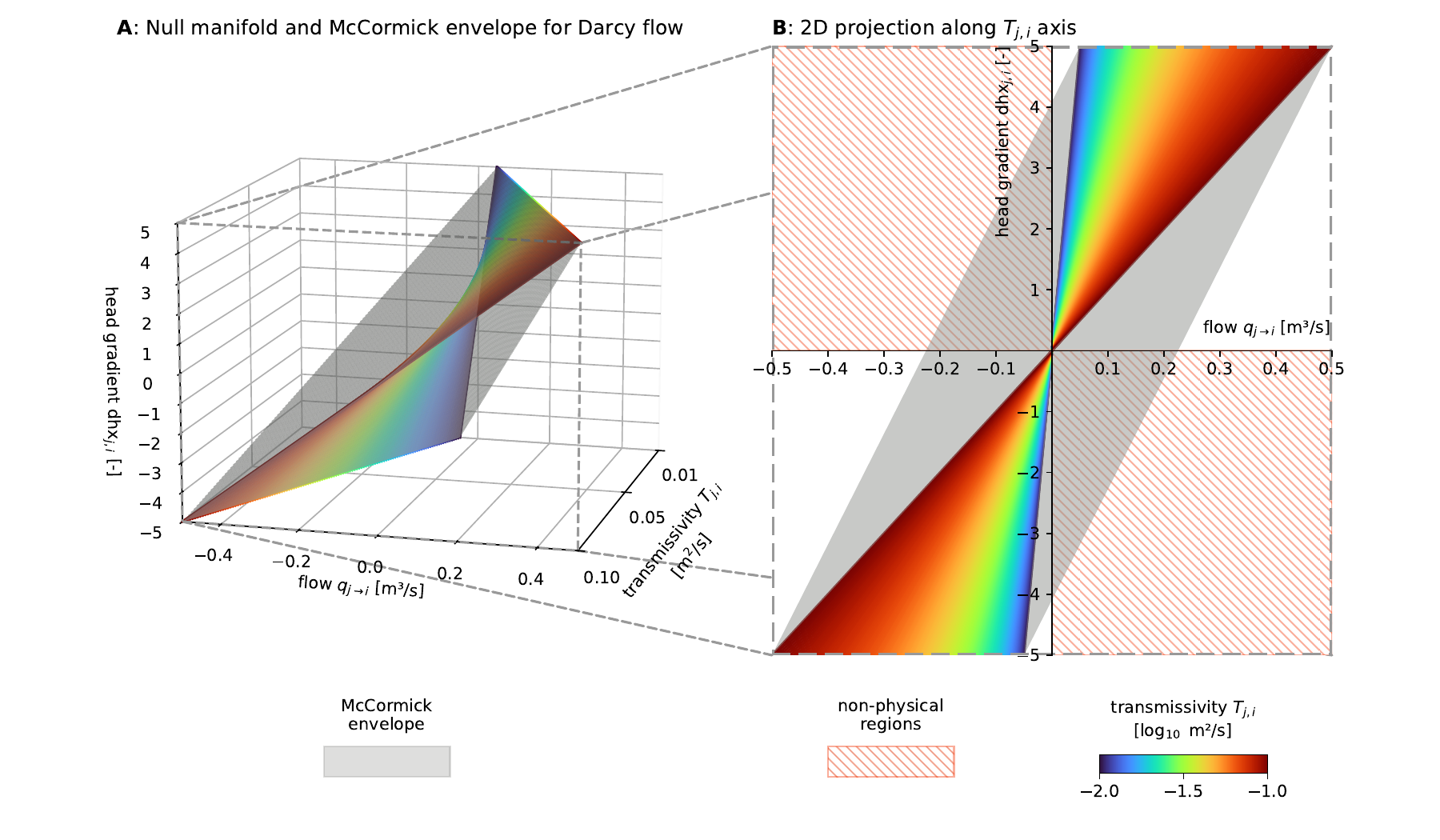}
  \caption{(A) The null manifold (colored surface) of the bilinear product $Q_{j\rightarrow i} = T_{j,i} \operatorname{dhx}_{j\rightarrow i}$ and its McCormick relaxation (grey) 3D. (B) 2D projection of the manifold and McCormick envelope along the $T_{j,i}$ axis. The McCormick envelope extends into non-physical quadrants.}
  \label{fig:McCormick}
\end{figure}

Figure~\ref{fig:McCormick}A shows the true nonlinear and nonconvex null manifold (colored surface) and the surrounding linear and convex McCormick envelope (grey). In subplot~\ref{fig:McCormick}B, we see that same solution manifold as a 2D projection along the transmissivity axis. From this perspective, it becomes easier to recognize the physics: the projected null manifold comprises of two triangular feasible regions touching at the origin. This double-triangular shape should make sense intuitively: for high transmissivities, we can achieve high flow magnitudes with small head gradients; lower transmissivities demand larger gradients to achieve the same flow magnitude. The point-symmetric shape around the origin arises because the flow and head-gradient directions can be negative- or positive-valued.

Note that the projected McCormick envelope in Figure~\ref{fig:McCormick}B relaxes the feasible region to include regions of variable space in the non-physical quadrants in the top-left and the bottom-right. These quadrants are non-physical because they imply positive flow induced by negative head gradients, or vice versa. We will revisit this challenge in Section~\ref{subsec:2D-steady-state}.

\subsection{Pseudo-code}

With the mechanisms of OBBT established, we can introduce the pseudo-code (Algorithm~\ref{alg:OBBT}) for the OBBT algorithm employed in the following. The code is based on a graph $\mathcal{G}$ of the model's spatial grid, with different variables defined along either nodes or edges of this grid. Transient simulations are likewise represented as a graph, the spatial grid "stacked" once for each timestep with edges connecting the same cells across adjacent timesteps, resulting in a spatio-temporal grid.
\begin{algorithm}[!h]
\SetAlgoLined
\DontPrintSemicolon

\BlankLine
\textbf{Input:} grid graph $\mathcal{G}$, initial bounds $(\underline{x},\overline{x})$ for $\{h_i^t, \; R_i^t, \; S_i, \; T_{j,i}\}$, tolerance $\varepsilon$, max iterations $K$\;
\textbf{Output:} tightened bounds $(\underline{x},\overline{x})$\;

\BlankLine
\textbf{Build }$\operatorname{variables}${ on }$\mathcal{G}$:\;
\begin{itemize}
    \item Nodes $(t,i)$: Initialize $h_i^t, \; R_i^t, \; S_i$
    \item Spatial edges $(t,j)\rightarrow(t,i)$: Initialize $T_{j,i}, \; \operatorname{dhx}_{j\rightarrow i}^{t}, \; q_{j\rightarrow i}^{t}$
    \item Temporal edges $(t-1,i)\rightarrow(t,i)$: Initialize $\operatorname{dht}_{i}^{t}, \; q_{i}^{t-1\rightarrow t}$
\end{itemize}
Initialize bounds $(\underline{x},\overline{x})$ for $h_i^t, \; R_i^t, \; S_i, \; T_{j,i}$\;
Derive initial bounds for flows $q$, $\operatorname{dhx}$, $\operatorname{dht}$ via interval arithmetic (Equations~\ref{eq:Darcys_law}~\&~\ref{eq:head_gradient})\;

\BlankLine
\textbf{Construct base constraints } $\mathcal{C}_{\text{base}}$:\;
\Indp
Mass balance equalities; observation bound constraints; other linear constraints\;
\Indm

\BlankLine
\For{$k \leftarrow 1$ \KwTo $K$}{

  \BlankLine
  Build/Update McCormick relaxations $\mathcal{C}_{\text{Mc}}(\underline{x},\overline{x})$\;

  \BlankLine
  Define LP feasible set $\mathcal{C} \leftarrow \mathcal{C}_{\text{base}} \cup \mathcal{C}_{\text{Mc}}(\underline{x},\overline{x})$\;

  \BlankLine
  Create a copy of the current bounds
  $(\underline{x}',\overline{x}') \leftarrow (\underline{x},\overline{x})$\;

  \BlankLine
  \textbf{Extremize via LP}:\;
  \ForEach{$x \in \operatorname{variables}$}{
    $\underline{x}' \leftarrow \max\{\underline{x},\ \min\{x \mid \mathcal{C}\}\}$\;
    $\overline{x}' \leftarrow \min\{\overline{x},\ \max\{x \mid \mathcal{C}\}\}$\;
  }

  \BlankLine
  Tighten bounds $(\underline{x},\overline{x}) \leftarrow (\underline{x}',\overline{x}')$\;

  \BlankLine
  Tighten bounds of bilinear products (Equation~\ref{eq:Darcys_law}) via interval arithmetic

  \BlankLine
  Stop if converged:\;
  \If{$1 - \prod_{x_i \in \operatorname{variables}} (\overline{x_i}'-\underline{x_i}')/(\overline{x_i}-\underline{x_i}) \le \varepsilon $}{
    \textbf{break}\;
  }
}

\caption{Optimization-based bound tightening (OBBT) with McCormick relaxations}
\label{alg:OBBT}
\end{algorithm}

Because the McCormick relaxations are constructed with a variable's current bounds $x \in [\underline{x},\overline{x}]$, the algorithm is iterative: after each of the $K \in \mathbb{N}$ iterations, new McCormick constraints are constructed. This repeats until either the $K$th iteration has been reached, or the bound tightening falls below a specified threshold.

In practice, we observed that the McCormick relaxations often do not sufficiently propagate information from the flows $q_{j \rightarrow i}^{t}$ and $q_{i}^{t-1 \rightarrow t}$ to the gradients $\operatorname{dhx}_{j \rightarrow i}^{t}$ and $\operatorname{dht}_{i}^{t}$. For instance, the flows may become sign-definite but the gradients may remain sign-ambiguous. To rectify this, we add a small post-processing step after each bound tightening, in which we use interval arithmetic \citep[e.g., ][]{moore2009introduction} to tighten the bounds of the bilinear products (Equation~\ref{eq:Darcys_law}). This ensures sign consistency between flows and gradients.

Note that while the computational demand for OBBT can be substantial compared to many sample-based UQ methods, the code is embarrassingly parallel. While the OBBT iterations must occur in sequence, the individual extremization operations of the optimization loop (i.e., "$\textbf{foreach }x\in\operatorname{variables}\textbf{ do}$" in Algorithm~\ref{alg:OBBT}) -- in which the algorithm spends most of its time -- can be solved in parallel.

\section{Simulations and Results}\label{sec:examples}

With the methodological basics established, we will explore the application of the OBBT framework to hydrogeological uncertainty estimation in three numerical example cases:
\begin{enumerate}
    \item A \textbf{1D steady-state} groundwater model to demonstrate the basic principles of hydrogeological OBBT in a simple setting that permits intuitive explanations about the resulting bounds, including a comparison to sample-based methods.
    \item A \textbf{2D steady-state} groundwater model to highlight challenges arising from the McCormick relaxations, and options on how to resolve them.
    \item A \textbf{2D transient} groundwater model to illustrate the application of OBBT for time-varying systems, making use of advanced hydrogeological constraints.
\end{enumerate}

The OBBT implementation used in the following leverages the $\operatorname{linprog}$ function of $\operatorname{scipy}$ \citep{virtanen2020scipy} in Python, which provides an interface to the HiGHS \citep{huangfu2018parallelizing} interior-point solver we use for the LP solves.

\subsection{1D steady-state}\label{subsec:1D-steady-state}

\subsubsection{Model setup}

We begin with a 1D steady-state groundwater model with 10 cells of width $dx=w=10$~m. The left-most cell contains an injection well with rate $R_{1}\in[10^{-6},10^{-4}]$~[m/s], and the right-most cell contains an extraction well with rate $R_{10}\in[-10^{-3},-10^{-5}]$~[m/s]. Recharge is zero in all intermediate cells. Transmissivities are independent and bounded by $T_{j,i}\in[10^{-4},10^{-1}]$~[m$^2$/s]. Initial hydraulic-head bounds are $h_i\in[3,12]$~[m] for all cells. We assume perfect head observations in cells 4 and 7, which fix these values at $h_4=10$~[m] and $h_7=7.0$~[m].

\subsubsection{Empirical verification}

We compare the OBBT bounds against two sample-based references. The first is an exact Monte Carlo sampler that exploits the simplicity of this 1D model to draw samples directly from the true null manifold. Its construction is described in Appendix~\ref{appendix:exact_sampler}. Figure~\ref{fig:experiment_1D}A--D shows $N=1000$ such samples as orange lines.

The second reference is a Markov chain Monte Carlo (MCMC) sampler representing a conventional Bayesian workflow. We use an Adaptive Differential Evolution Metropolis step \citep{ter2008differential} as implemented in PyMC \citep{pymc2023}, with four chains, $10{,}000$ tuning steps, and up to $25{,}000$ sampling steps per chain, repeated over ten random seeds. For smaller sample sizes, we truncate the chains earlier.

We place a uniform prior over the initial variable bounds and use a Gaussian likelihood. For each run, we estimate the marginal minimum and maximum values of the posterior samples for four variable types: $T_{j,i}$ (log scale), $h_i$ (linear scale), $R_i$ (signed log scale), and $q_{j\rightarrow i}$ (log scale). We then average these bounds across random seeds. This lets us quantify how much of the null manifold, or equivalently the MAP ridge in this non-identifiable problem, is explored by chains of different lengths.

OBBT, by contrast, solves the exact-fit feasibility problem. In Bayesian terms, this corresponds to a Dirac delta likelihood, or a Gaussian likelihood with $\sigma_{\mathrm{obs}}=0$. An MCMC sampler cannot explore that set directly, because a random proposal lands on the exact null manifold with probability zero. We therefore relax the comparison by using three nonzero observation errors, $\sigma_{\mathrm{obs}}\in\{0.001,0.01,0.1\}$~[m]. This makes the comparison slightly favorable to MCMC, because larger $\sigma_{\mathrm{obs}}$ admit increasing slack around the observations.

\subsubsection{Results}

\begin{figure}
  \centering
  \includegraphics[width=\textwidth]{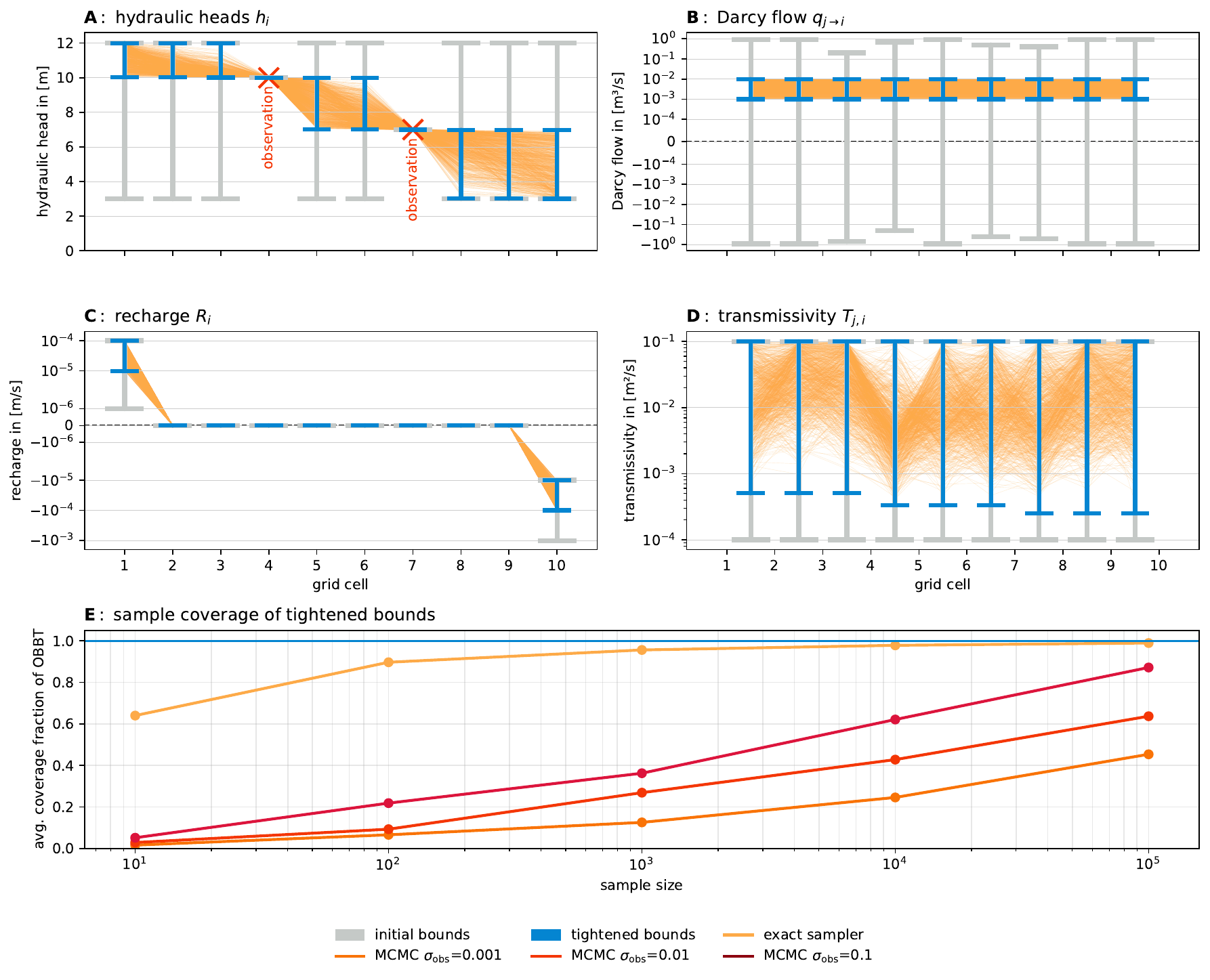}
  \caption{Initial (grey) and tightened (blue) interval bounds for hydraulic head (A), Darcy flow (B), recharge (C), and transmissivity (D). Heads and recharge are defined in cells; transmissivity and Darcy flow are defined between cells. Orange lines show $1000$ exact samples fitted to the observations within the initial bounds. Panel~E shows the average marginal coverage of the OBBT bounds for $T_{j,i}$, $h_i$, $R_i$, and $q_{j\rightarrow i}$ by samples from the exact null manifold sampler and from three MCMC chains with different values of $\sigma_{\mathrm{obs}}$}
  \label{fig:experiment_1D}
\end{figure}

Figure~\ref{fig:experiment_1D}A shows the initial and tightened bounds for hydraulic head. Both the lower bounds $\underbar{h}$ and the upper bounds $\bar{h}$ decrease monotonically from left to right, consistent with the source--sink configuration. The intervals narrow sharply around the observations in cells 4 and 7, which in turn reduces the admissible head range in neighboring cells.

Initial flow bounds are sign-ambiguous (Figure~\ref{fig:experiment_1D}B), so the algorithm initially does not know the direction of flow between adjacent cells. Because these bounds are derived from the initial bounds on $T_{j,i}$ and $\operatorname{dhx}_{j\rightarrow i}$, the initial bounds on $q_{j\rightarrow i}$ are already somewhat narrower near observed cells. During tightening, OBBT identifies the physically consistent flow sign and contracts the intervals accordingly.

Initial recharge bounds differ in magnitude between the source and sink cells (Figure~\ref{fig:experiment_1D}C). The first tightening step enforces mass balance: under steady-state conditions, total injection cannot exceed total extraction, and total extraction cannot exceed total injection.

The transmissivity bounds (Figure~\ref{fig:experiment_1D}D) separate into three regions: left of the first observation, between the two observations, and right of the second observation. In all three regions, OBBT raises only the lower bounds $\underbar{T}$. This is expected. The observations and initial head bounds limit the admissible head drop to at most $2$~m on the left, exactly $3$~m in the center, and at most $4$~m on the right. Small values of $T_{j,i}$ can require head drops larger than these limits and are therefore excluded. Large transmissivities remain feasible because they can be offset by smaller transmissivities elsewhere in the same region. Appendix~\ref{appendix:exact_sampler} examines this effect in more detail.

The exact Monte Carlo samples support these results: none fall outside the OBBT bounds. The OBBT loop requires only two iterations and $1.2$~s to reduce the product of marginal bound shrinkages below $0.1$~\%. This corresponds to about $6.5$~ms per single minimization or maximization.

Figure~\ref{fig:experiment_1D}E compares the average marginal coverage of the OBBT bounds by the exact sampler and by the three MCMC configurations. Coverage increases with sample size in all cases, but every sample-based method underestimates the extent of the feasible region. The exact sampler converges toward the OBBT bounds, yet finite samples still only provide underestimates. 

MCMC performs substantially worse. Acceptance rates range from $3.8$~\% for $\sigma_{\mathrm{obs}}=0.001$ to $8.3$~\% for $\sigma_{\mathrm{obs}}=0.1$. Even with $100{,}000$ samples, the most permissive chain reaches only about $87$~\% coverage, despite solving a more forgiving problem than OBBT. The chain closest to the exact-fit setting, with $\sigma_{\mathrm{obs}}=0.001$, covers only about $45$~\% of the marginal admissible range on average.

This experiment supports three conclusions. First, OBBT yields physically plausible marginal bounds that enclose the null manifold of this non-identifiable groundwater model. Second, in Bayesian terms, these bounds enclose the MAP ridge induced by a non-informative prior. Third, finite sample-based methods can severely underestimate the extent of that ridge, even in a simple 1D problem. Due to the curse of dimensionality, we expect this phenomenon to become much worse in more complex models, where the number of parameters will be orders of magnitude larger and the sample sizes orders of magnitude smaller.

\subsection{2D steady-state}\label{subsec:2D-steady-state}

This experiment extends the 1D example to two dimensions and exposes the main pathology of hydrogeological OBBT under McCormick relaxations. We consider four variants of the same model: a pure-LP reference solution, a basic OBBT formulation, OBBT with irrotationality constraints, and OBBT with prescribed flow signs. Together, these cases show why naive bound tightening fails and which additional constraints restore informative bounds.

\subsubsection{Model setup}

\begin{figure}
  \centering
  \includegraphics[width=\textwidth]{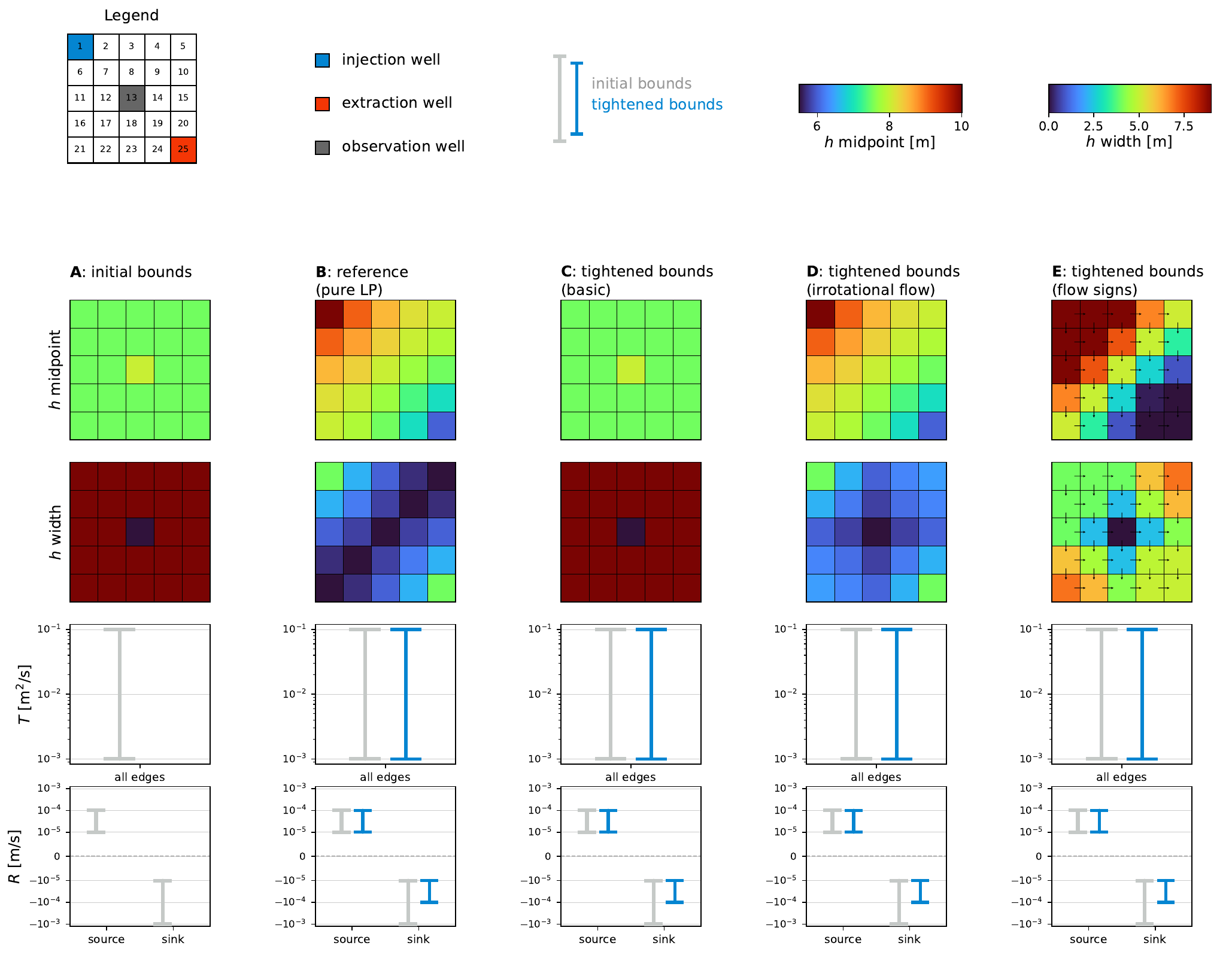}
  \caption{Bound-tightening results for the 2D steady-state example. Panel~A shows the initial bounds for $h_i$, $T_{j,i}$, and $R_i$. Panel~B shows the pure-LP reference solution. Panels~C--E show the tightened bounds obtained from basic OBBT, OBBT with irrotationality constraints, and OBBT with prescribed flow signs, respectively. Intervals for distributed variables ($h_i$) are shown as spatial maps; intervals for scalar quantities are shown as 1D spans.}
  \label{fig:experiment_2D_steady_state}
\end{figure}

We use a $5\times 5$ grid of rectangular cells with spacing $dx=10$~m. The top-left cell contains a source with recharge $R_{1}\in[10^{-5},10^{-4}]$~[m/s], and the bottom-right cell contains a sink with recharge $R_{25}\in[-10^{-3},-10^{-5}]$~[m/s]. A perfect hydraulic-head observation is prescribed in the center cell, $h_{13}=8$~[m]. Transmissivity is homogeneous and bounded by $T_{j,i}=T\in[10^{-3},10^{-1}]$~[m$^2$/s]. Initial hydraulic-head bounds are $h_i\in[3,12]$~[m]. The legend of Figure~\ref{fig:experiment_2D_steady_state} summarizes the setup.

\subsubsection{Experiment A: Reference solution}

We first construct a reference solution. In this homogeneous setting, we can avoid McCormick relaxations by fixing $T$ to constant values and solving a sequence of pure LPs. Specifically, we run 101 OBBT passes with $T$ fixed at evenly spaced values on a log scale over $[10^{-3},10^{-1}]$~[m$^2$/s]. With $T$ fixed, all bilinear terms become linear. We then take the union of the tightened bounds across all transmissivity slices.

Figure~\ref{fig:experiment_2D_steady_state}B shows the result. Transmissivity does not tighten, so the full prescribed range of $T$ remains feasible. The sink recharge $R_{25}$ tightens to balance the source. The hydraulic-head bounds show the expected gradient from source to sink. The central observation $h_{13}=8$~[m] acts as a pivot, collapsing uncertainty along the northeast-southwest axis, orthogonal to the source-sink dipole. The bounds widen again toward the source and sink because different combinations of $T$ and $R$ still permit steeper or flatter head fields that remain consistent with the central observation.

\subsubsection{Experiment B: Basic OBBT}

Figure~\ref{fig:experiment_2D_steady_state}C shows the result of basic OBBT, that is, OBBT without any additional constraints. The outcome is striking: neither $h_i$ nor $T$ tightens. Only the sink recharge $R_{25}$ contracts, again to balance the source $R_1$.

The reason is that mass balance alone is not enough. Recharge variables are constrained directly by the linear mass-balance equations, so OBBT tightens them exactly. Hydraulic heads, however, depend on how flux information propagates through Darcy's law. In the relaxed problem, the signs of the volumetric fluxes $q_{j\rightarrow i}$ remain ambiguous everywhere, including along edges adjacent to the source and sink. This is possible because the McCormick relaxation admits non-physical combinations of flux and head gradient in which flow can occur against the hydraulic-head gradient (Figure~\ref{fig:McCormick}B). The relaxation therefore breaks the sign coupling between $q_{j\rightarrow i}$ and $\operatorname{dhx}_{j\rightarrow i}$.

\begin{figure}
  \centering
  \includegraphics[width=\textwidth]{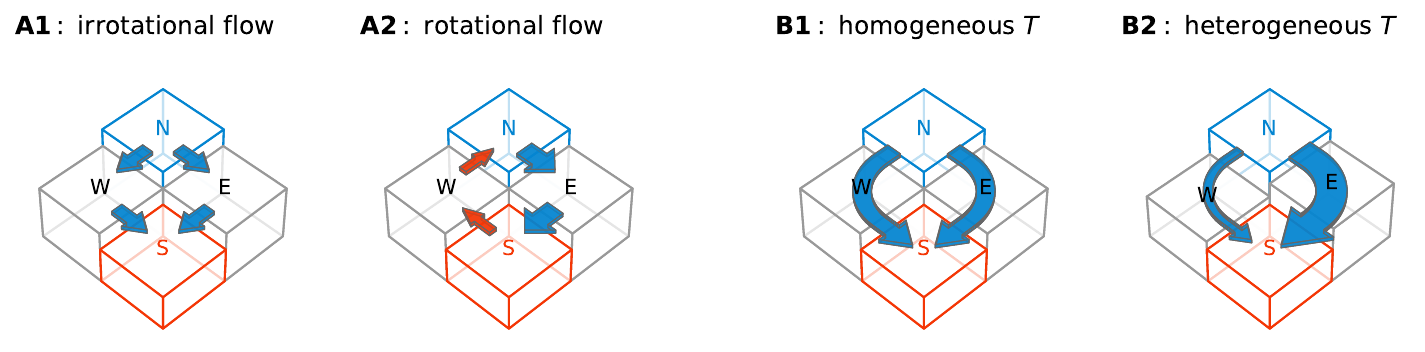}
  \caption{Conceptual illustration of rotational-flow pathology and its remedies on a 2-by-2 grid with source N and sink S. Arrow width indicates flow magnitude. Under Darcy flow, transport is irrotational and follows the head gradient (A1). The relaxed system can also admit a circulating flow pattern that satisfies mass balance but is non-physical under Darcy flow (A2). In homogeneous media, the two source-to-sink paths carry equal flux, so zero signed cycle flow is a valid irrotationality constraint (B1). In heterogeneous media, the flow may remain irrotational while the two path fluxes differ, so this constraint no longer applies in general (B2).}
  \label{fig:irrotational_flow}
\end{figure}

That sign coupling is essential because it rules out rotational flow. To see this, consider a $2\times2$ grid with cells $N$, $E$, $S$, and $W$, where $N$ is a source and $S$ is a sink. Irrotational flow would move water from $N$ to $S$ along the two paths $N\rightarrow E\rightarrow S$ and $N\rightarrow W\rightarrow S$ (Figure~\ref{fig:irrotational_flow}A1). Mass balance alone, however, also permits a cycle (Figure~\ref{fig:irrotational_flow}A2)
\begin{equation}
    N \xrightarrow[]{q_{N\rightarrow E}\geq 0} E \xrightarrow[]{q_{E\rightarrow S}\geq 0} S \xrightarrow[]{q_{S\rightarrow W}\geq 0} W \xrightarrow[]{q_{W\rightarrow N}\geq 0} N.
    \label{eq:flow_cycle}
\end{equation}
Under exact Darcy flow, such a cycle is impossible. If each edge carries nonnegative flow in the indicated direction, Darcy's law implies
\begin{equation}
    h_{N} \geq h_{E} \geq h_{S} \geq h_{W} \geq h_{N}.
    \label{eq:flow_cycle_inequality}
\end{equation}
Because $h_N$ appears at both ends of this chain, all inequalities must be equalities:
\begin{equation}
    h_{N} = h_{E} = h_{S} = h_{W} = h_{N}.
\end{equation}
The head gradients along the cycle must therefore vanish,
\begin{equation}
     \operatorname{dhx}_{N\rightarrow E} = \operatorname{dhx}_{E\rightarrow S} = \operatorname{dhx}_{S\rightarrow W} = \operatorname{dhx}_{W\rightarrow N} = 0,
\end{equation}
and so must the corresponding flows,
\begin{equation}
     q_{N\rightarrow E} = q_{E\rightarrow S} = q_{S\rightarrow W} = q_{W\rightarrow N} = 0.
\end{equation}
The McCormick relaxation breaks this implication: Equation~\ref{eq:flow_cycle} no longer enforces Equation~\ref{eq:flow_cycle_inequality}. Rotational flow becomes feasible, and once it does, OBBT can no longer infer informative bounds on $h_i$ from $q_{j\rightarrow i}$, or vice versa.

\subsubsection{Experiment C: OBBT with irrotationality constraints}\label{subsubsec:2D_irrotationality}

To test this diagnosis, we add explicit irrotationality constraints. In a steady-state, homogeneous, isotropic system on a regular grid, irrotationality implies that the directed flux around any cycle in the grid graph $G$ must sum to zero:
\begin{equation}
    \sum_{(j\rightarrow i)\in \operatorname{cycle}} q_{j\rightarrow i} = 0
    \qquad \forall \ \operatorname{cycle}\in C(G),
    \label{eq:irrotationality_constraint}
\end{equation}
where $C(G)$ denotes the fundamental cycle basis of $G$. We therefore identify all fundamental cycles and add Equation~\ref{eq:irrotationality_constraint} as an equality constraint for each of them.

Figure~\ref{fig:experiment_2D_steady_state}D shows the result. After the first OBBT iteration, the system resolves the flow directions throughout the grid. The midpoint of the tightened $h_i$ bounds matches that of the reference solution in Figure~\ref{fig:experiment_2D_steady_state}B, while the intervals remain slightly wider in the northeastern and southwestern corners. This is expected: the formulation is still a relaxation and therefore need not recover the exact reference bounds.

\subsubsection{Experiment D: OBBT with prescribed flow signs}\label{subsubsec:2D_flow_sign}

The irrotationality constraints of Section~\ref{subsubsec:2D_irrotationality} are useful in this simple setting, but they are not generally applicable. If $T_{j,i}$ is heterogeneous or anisotropic, zero net directed flux around a cycle no longer follows from irrotationality (Figure~\ref{fig:irrotational_flow}B1 \& B2). We therefore consider a weaker but more general alternative: prescribing the sign of the flow across each interface $(j,i)$. This removes one wedge of the McCormick envelope in Figure~\ref{fig:McCormick} and substantially tightens the relaxation.

Figure~\ref{fig:experiment_2D_steady_state}E shows the resulting bounds. The midpoint of $h_i$ again traces a gradient from source to sink, although it differs from the midpoint in both the reference case and the irrotationality-constrained case. This is not problematic: the midpoint is only a summary of the interval centers and need not itself correspond to a feasible realization.

The resulting $h_i$ bounds are wider than those in Section~\ref{subsubsec:2D_irrotationality}. In this 2D example, prescribed flow signs are weaker than explicit irrotationality constraints. In this homogeneous setting, the irrotationality constraint in Experiment~C not only fixes flow direction indirectly, but also constrains flow magnitude, because it implies that $q_{j\rightarrow i}$ must match along any two non-intersecting paths between source and sink.

\subsection{2D transient}\label{subsec:2D-transient}

\subsubsection{Model setup}

Our final example is a 2D transient groundwater model on a regular hexagonal grid with cell spacing $dx=10$~m. A small no-flow region is removed from the western part of the domain. We simulate five timesteps of length $\Delta t = 1$~[d].

The northwestern part of the domain contains a recharge zone consisting of seven cells. In each of these cells, recharge is bounded by $R_i^{+}\in[10^{-6},10^{-5}]$~[m/s] during the first three timesteps and fixed to zero thereafter. An extraction well in the northeast becomes active at $t=2$ and is bounded by $R_{\text{well}}^{-}\in[-10^{-3},-10^{-5}]$~[m/s]. This pumping rate is represented by a single shared variable across all active timesteps. Along the southern boundary, hydraulic head is prescribed by a single fixed value $h^{\text{boundary}}=7.0$~[m], shared across all boundary cells and timesteps.

Initial head bounds in all other cells are $h_i\in[0.0,12.0]$~[m]. Specific yield is cell-specific and bounded by $S_i\in[0.15,0.25]$~[-]. Transmissivities are bounded by $T_{j,i}\in[10^{-4},10^{-2}]$~[m$^2$/s] and are constant through time. To impose weak local smoothness, we constrain transmissivities incident to the same cell to remain within a factor $1+\rho$ of one another. That is, for any two transmissivities $T_{j,i}$ and $T_{k,i}$ connected to the same cell $i$, we impose
\[
T_{j,i}\leq (1+\rho)\,T_{k,i}
\qquad \text{and} \qquad
T_{k,i}\leq (1+\rho)\,T_{j,i},
\]
with $\rho=0.05$.

As in Section~\ref{subsubsec:2D_flow_sign}, we prescribe signs to preserve the coupling between gradients and flows. To obtain a physically plausible sign pattern, we first compute a midpoint FVM reference simulation using midpoint values of $T_{j,i}$, $S_i$, $R_i^t$, and $h^{\text{boundary}}$. We then extract both the spatial flow directions and the temporal signs of head change and impose them as sign constraints in the OBBT formulation. From the same reference simulation, we extract hydraulic heads at three observation cells and impose them as synthetic observation constraints at all timesteps.

\subsubsection{Results}

\begin{figure}
  \centering
  \includegraphics[width=\textwidth]{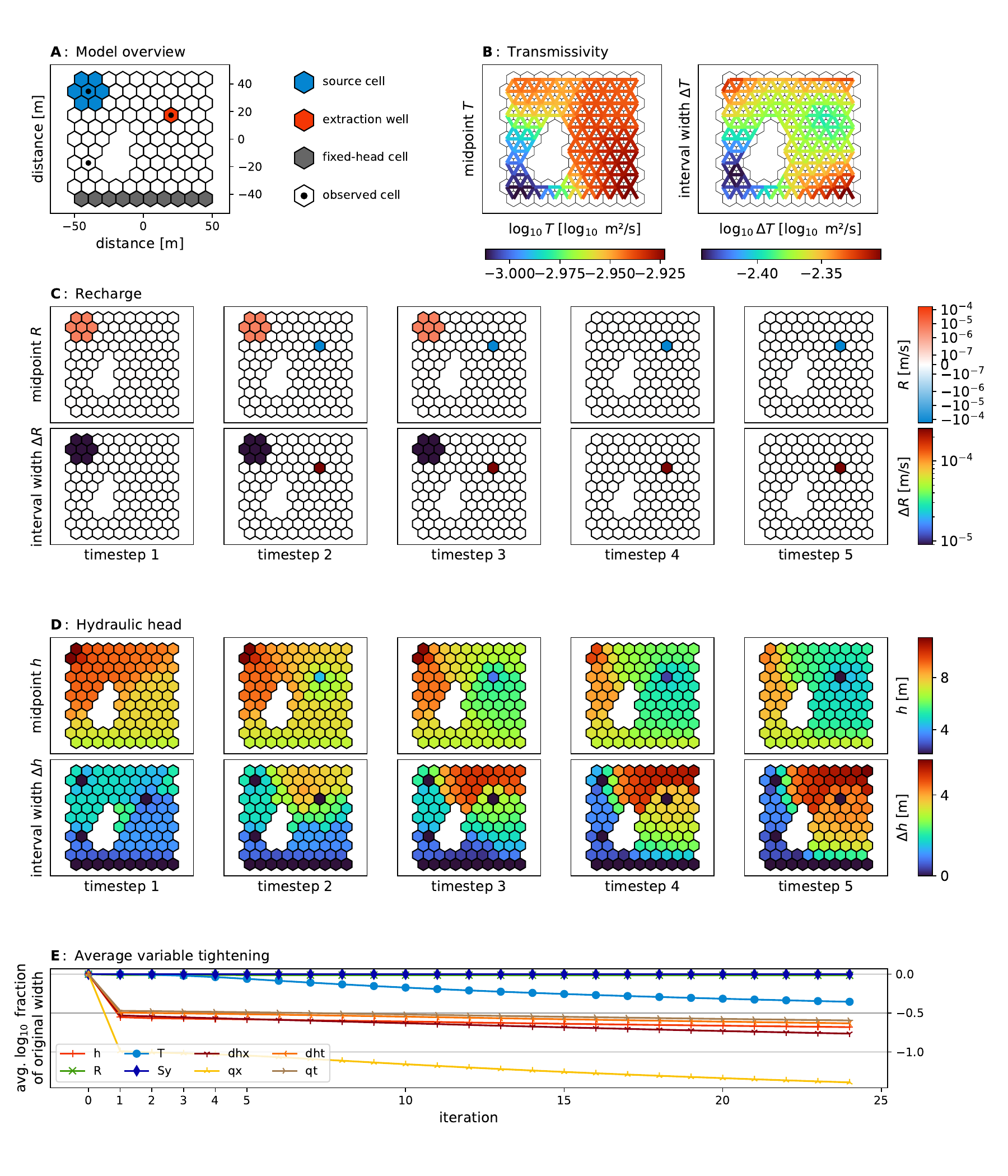}
  \caption{Model setup (A), tightened bounds for transmissivity $T_i$ (B), recharge $R_i^t$ (C), and hydraulic head $h_i^t$ (D), and average tightening per OBBT iteration and variable type (E), computed as $1-(\overline{x}'-\underline{x}')/(\overline{x}-\underline{x})$.}
  \label{fig:experiment_2D_transient}
\end{figure}

Figure~\ref{fig:experiment_2D_transient} summarizes the tightened bounds. Panel~A shows the model setup. Panels~B--D show, for transmissivity, recharge, and hydraulic head, both the midpoint of the tightened intervals and their remaining width.

Transmissivity tightens only slightly (Figure~\ref{fig:experiment_2D_transient}B). The midpoint field remains close to the initial value. The interval width decreases most significantly near the extraction well and in the South-West, the narrow region near the boundary-adjacent observation well. Note that both areas are where the flow is most constrained. The seven recharge cells in the northwest retain their initial bounds during the first three timesteps, whereas the extraction well tightens noticeably once pumping begins. Specific yield is not shown because its bounds do not tighten (see Figure~\ref{fig:experiment_2D_transient}E).

Hydraulic head shows the strongest response (Figure~\ref{fig:experiment_2D_transient}D). At $t=1$, before the extraction well becomes active, the midpoint field reflects a quasi-steady pattern controlled by recharge in the northwest and the fixed-head boundary in the south. Once pumping begins at $t=2$, a drawdown cone develops around the well and expands through time. This effect becomes stronger after the recharge zone turns off at $t=4$. The interval widths contract most strongly in the northern and central parts of the domain, while remaining comparatively narrow near the fixed-head boundary where head is already constrained.

Figure~\ref{fig:experiment_2D_transient}E shows the average remaining interval width per iteration for $h_i^t$, $R_i^t$, $T_{j,i}$, $S_i$, $dhx_{j\rightarrow i}^t$, $q_{j\rightarrow i}^t$, $dht_i^t$, and $qt_i^t$. The most significant tightening occurs in the first iteration. Subsequent iterations tighten variables by iteratively refining the McCormick relaxations, so that $T_{j,i}$ only starts to tighten appreciably in later iterations.

The full run required $71.7$~h of wall-clock time on sixteen CPU cores, or about $1.38$~s per variable extremization. This cost is substantial, but the output is also different in kind from that of a forward simulation or a Monte Carlo ensemble. OBBT does not produce one realization or a finite ensemble; it produces outer bounds on all uncertain variables simultaneously. In that sense, the cost is the price of conservative coverage rather than pointwise exploration.

Even so, the run could likely be shortened considerably. Figure~\ref{fig:experiment_2D_transient}E suggests that little is gained after the first five iterations, and later iterations could also be restricted to variables that tightened appreciably in the previous pass. We used neither shortcut here: all variables were tightened in all 25 iterations.

\section{Discussion \& Outlook}\label{sec:discussion}

\subsection{Discussion}\label{subsec:discussion}

In this study, we introduced optimization-based bound tightening (OBBT) as an interval-based framework for UQ in non-identifiable groundwater models. Rather than exploring uncertainty through finite samples or ensembles, we formulated prior bounds, observations, mass balance, and discretized Darcy flow as a constraint system and used repeated linear optimization to compute safe marginal bounds on hydraulic heads, fluxes, recharge, transmissivity, and storage-related variables. To make the uncertain Darcy-flow equations tractable, we combined a finite-volume discretization with McCormick relaxations of the resulting bilinear terms. Through steady-state and transient numerical examples, we showed that OBBT can enclose the admissible null manifold of groundwater models, but also that naive relaxations can admit non-physical flow behavior unless additional physical constraints, such as flow-sign prescriptions or irrotationality constraints, are imposed.

These results leave one central challenge: the usefulness of OBBT depends on how much physical structure survives the relaxation. The following points summarize where the method works well, where the relaxation becomes too permissive, and how additional constraints can recover informative bounds.
\begin{enumerate}
    \item \textbf{Fixing variable values} for either (a) head fields or (b) transmissivity and specific yield, which turns the bilinear constraints into linear constraints and circumvents the need for McCormick relaxations altogether (\textcolor{red}{strong assumption}, \textcolor{green_custom}{computationally cheap}).
    \item Prescribing (physically plausible) \textbf{flow signs} across each cell edge, which ensures irrotationality explicitly (\textcolor{orange}{moderately strong assumption}, \textcolor{orange}{computationally efficient}). 
    \item Formulating \textbf{piecewise McCormick relaxations} with a binary variable split at zero, which activates or deactivates one of the quadrants in figure~\ref{fig:McCormick}B. The result would be a mixed-integer LP that explores different flow sign combinations. We have not discussed such an approach in this study, as the resulting MILP may become computationally infeasible for even moderate grid sizes, but note that efficient coupling between the binary variables may ameliorate these issues. (\textcolor{green_custom}{weak assumption}, \textcolor{red}{computationally expensive}).
\end{enumerate}
While all three strategies are viable and could, in principle be combined freely, we focused on Option~2 in this study. While prescribing flow signs may sound like a strong assumption, we note that it is three levels more general than a conventional inverse problem. In increasing order of generality:
\begin{enumerate}[label=\alph*.]
    \item A classical, possibly nonidentifiable, inverse problem assumes knowledge of the model output, that is to say, the full \textbf{hydraulic head field} $h_i^t$. Knowledge of the head field implies knowledge of the head gradients $\operatorname{dhx}_{j\rightarrow i}^{t}$ and therefore resolves the bilinearity, resulting in a pure LP. 
    \item Knowing only the \textbf{hydraulic head gradient} $\operatorname{dhx}_{j\rightarrow i}^{t}$ likewise resolves the bilinearity and results in a pure LP, but offers an additive degree of freedom due to an unspecified offset.
    \item Prescribing only \textbf{flow directions} is a weaker assumption, as there may be infinitely many $h_i^t$ and $\operatorname{dhx}_{j\rightarrow i}^{t}$ fields that correspond to the same flow direction field if $T_{j,i}$ and $R_i$ are heterogeneous. Prescribing flow directions results in a bilinear constraint system that demands McCormick relaxation, but resolves the sign ambiguity.
    \item Prescribing only the \textbf{flow sign} is a weaker assumption still, as a flow sign field does not prescribe magnitude ratios between a cell's individual flow vectors. In more than one dimension, a single flow sign field can thus support infinitely many flow direction fields, which can in turn support infinitely many hydraulic head fields. Prescribing flow signs explicitly resolves the sign ambiguity.
\end{enumerate}

While OBBT is guaranteed to provide safe bounds under the user's prior assumptions, over-aggressive prior assumptions can still exclude otherwise viable solutions, just as in any other UQ method. It is important to acknowledge that prescribing flow signs constitutes a prior assumption. Incautious assignment of flow signs thus constitutes a potential risk for obtaining safe outer bounds on uncertain variables.

\begin{figure}
  \centering
  \includegraphics[width=\textwidth]{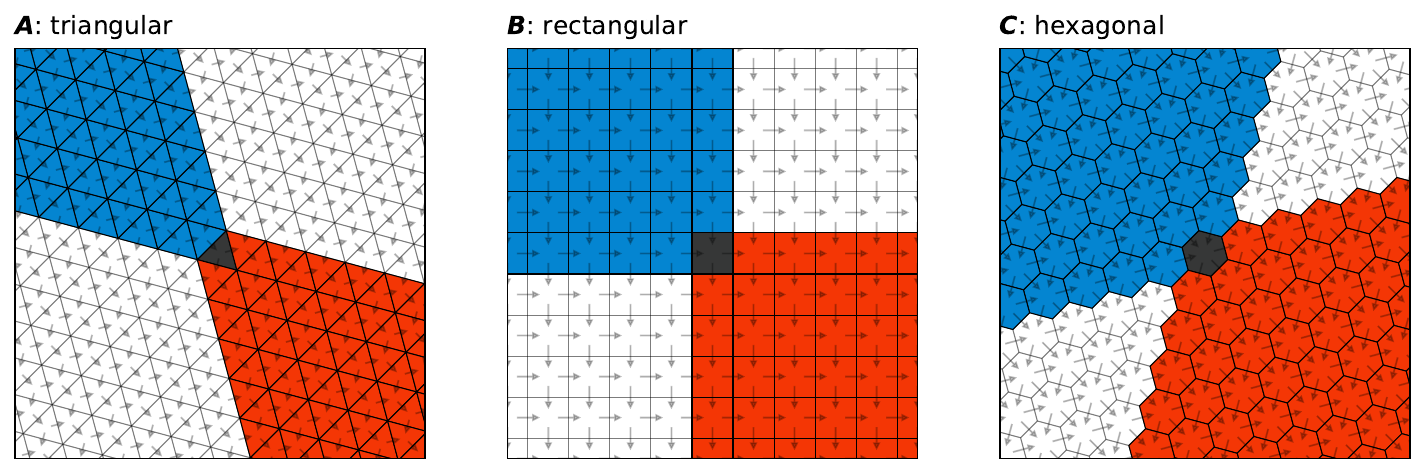}
  \caption{Three different regular grids, with a flow sign described from the top left to the bottom right. Relative to the gray center cell, cells marked in blue lie upstream and cells marked in red lie downstream. (A) A regular triangular grid permits the lowest degree flow direction uncertainty with a cone of 60°, (B) a regular square grid permits a cone of 90°, and a (C) hexagonal grid permits a cone of 120°.}
  \label{fig:flow_signs}
\end{figure}

We also note that grid geometry plays an important role in the permissibility of flow sign constraints. The more faces the cells of a regular grid have, the broader the range of potential flow directions it can represent after fixing the flow signs (Figure~\ref{fig:flow_signs}). This is because any specific flow direction is a composition of flow direction vectors of different magnitude, and the more diverse flow direction vectors are available, the greater the range of viable results. To cover the largest possible range of flow direction possibilities, we thus recommend using hexagonal grids.

Finally, it is important to acknowledge the high computational cost of the algorithm outlined in this study. OBBT in a discretized PDE-based system is more computationally demanding than conventional sample-based UQ, and it will likely remain so for the foreseeable future. Sample-based methods generally adjust ensemble size to the available computational resources, and so it is always possible to obtain \textit{some} empirical uncertainty estimate, no matter how high-dimensional or complex the underlying system. PDE-based OBBT, by contrast, is a young field, and the efficiency gains that transformed comparable numerical methods — finite element solvers, ensemble data assimilation — over successive decades of development have yet to materialize here. Likewise, the efficiency of LP solvers has increased significantly over the past decades, and further improvements will translate directly into efficiency gains of OBBT.

The more interesting question, in our opinion, is thus not one of computational cost, but of reliability. No sampling-based method, regardless of ensemble size, can provide a certified outer bound; it can only provide an empirical one. For decision support in water management, where regulators may need prove negatives, for instance to demonstrate that a proposed abstraction rate is safe under all admissible parameter combinations, this distinction is not merely academic. The tacit assumption that an empirical UQ estimate is sufficiently close to certified bounds is a tenuous proposition in the high-dimensional settings prevailing in hydrogeology. In decision support settings where the penalty for underestimating uncertainty is severe, the increased cost of OBBT is justified.

\subsection{Outlook}\label{subsec:outlook}

In this study, we provided a proof-of-concept for the use of OBBT for sample-free hydrogeological UQ. While initial results are promising, further research is required to operationalize the method at scales commonly encountered in hydrogeology, and improve the flexibility and efficiency of the framework. We want to conclude this manuscript by highlighting four interesting and complementary avenues for future research.

\paragraph{Generalized sign coupling} The greatest challenge of the approach outlined in this study is perhaps the weakened sign-coupling between the head gradients and the flows. In this study, we explored multiple options to ameliorate these issues. We proposed flow sign prescription as a versatile solution, but noted that explicit irrotationality constraints can yield tighter bounds in homogeneous transmissivity settings. Future research might focus on studying combinations of such constraints, for instance in zonal groundwater models, where irrotationality constraints could be applied inside each transmissivity zone. Further expanding this framework with an efficient MILP formulation for piecewise McCormick relaxations may result in more flexible and more powerful hydrogeological OBBT.

\paragraph{Interval-based data assimilation} A noteworthy side-effect of expressing the PDEs as a constraint system is that the physics are not enforced sequentially, but simultaneously across time. In consequence, constraints for future states (such as an observation at a later timestep) directly eliminate irreconcilable state values in the past. Conceptually, this shares similarity with \textit{Bayesian smoothers}, which likewise yield backwards-in-time consistency. One could envision a different, sequential-in-time OBBT architecture, in which we gradually construct the system of constraints one timestep at a time. Tightening only states and parameters at the latest timestep may result in more computationally efficient bound tightening at the expense of reduced backwards-in-time consistency; the resulting algorithm would behave similarly to a \textit{Bayesian filter}. Exploring these connections may not only yield improvements in OBBT efficiency, but would conceptually connect this framework to data assimilation.

\paragraph{System forensics} A PDE-based constraint system directly encodes the physical relationships. Analyzing the constraints can yield deeper insight into the physical system. For instance, evaluating the (relaxed) constraints with the feasible solutions identified during the OBBT routine can reveal which constraints are active at each extremum (geometrically: which faces of the polytope border this particular vertex). Doing the same for the \textit{original} nonlinear constraint system can either certify an extremum (thereby verifying this bound as optimal), or else provide insight into which constraints are violated, and by how much. Such analyses can aid the refinement of the OBBT and provide insight into physical relationships in the groundwater system.

\paragraph{Advanced hydrogeological constraints} In this study, we created a simple OBBT implementation with its physical constraints mostly based on Darcy flow PDEs. Future research could explore advanced constraint types, which may -- for instance -- leverage the fact that extrema in the $h_i^t$ field must always be located at source or sink cells. Likewise, it may be interesting to explore hydrogeological constraints that are difficult to enforce in other UQ approaches, such as Lipschitz bounds on $h_i^t$ through tighter initial constraints on $\operatorname{dhx}_{j\rightarrow i}^t$ and $\operatorname{dht}_i^t$. 

To conclude, we believe that hydrogeological OBBT is a very promising avenue for future research for UQ because it can, perhaps uniquely, provide safe outer bounds on uncertain variables.

\section*{Code availability}

The code to reproduce the figures and experiments in this manuscript is available under \url{https://github.com/MaxRamgraber/Bounding-The-Null-Space}.

\bibliographystyle{apalike}

\bibliography{references}

\appendix
\newpage
\section{Exact empirical sampler for the 1D steady-state simulation}\label{appendix:exact_sampler}

In this section, we describe the setup of the empirical sampler that generates exact Monte Carlo samples from the feasible region of the 1D steady-state experiment illustrated in Section~\ref{subsec:1D-steady-state}. Because this setting is conceptually very simple, it is possible to draw exact samples from the true feasible region.

To this end, we first split the system into three regions: left of the first observation $h_4$, between the observations $h_4$ and $h_7$, and right of the observation $h_7$. Based on initial bounds and the grid geometry, we compute the admissible total head drop in all three domain parts:
\begin{enumerate}
    \item Left: at most $2$~[m] head drop
    \item Center: exactly $3$~[m] head drop
    \item Right: at most $4$~[m] head drop
\end{enumerate}

\subsection{Sampling $q_{j \rightarrow i}$}

We then derive analytical bounds for the flow rate $q_{j \rightarrow i}$~[m³/s]. Because we only have a source ($R_1$) and sink ($R_{10}$) pair at the ends of the domain, the flow rate is constant throughout the domain. Naively, the bounds on the (absolute) volumetric flow rate $\underline{q_{j\rightarrow i}} \leq q_{j\rightarrow i} \leq \overline{q_{j\rightarrow i}}$~[m³/s] must be based on the recharge bounds:
$$
 \underline{q_{j\rightarrow i}} = \max(A\cdot \underline{R_1}, - A \cdot \overline{R_{10}}), \quad
\overline{q_{j\rightarrow i}} = \min(A\cdot \overline{R_1}, - A \cdot \underline{R_{10}})
$$

We can further tighten these flow bounds based on the admissible head drop over the three regions. For any collection of edges $(j,i) \in \operatorname{region}$ with a prescribed head drop $\Delta h_{\operatorname{region}}$, we can derive the maximum and minimum flow rate via Darcy's Law:

$$
q_{j\rightarrow i} = T_{j,i} \cdot w \cdot \frac{\Delta h_{\operatorname{j\rightarrow i}}}{\Delta x}
$$

With a bit of reformulation, for a given $q_{j\rightarrow i}$, we can equivalently compute what head drop $\Delta h_{\operatorname{j\rightarrow i}}$ is possible over that edge:
$$
\Delta h_{\operatorname{j\rightarrow i}} =   \cdot \frac{q_{j\rightarrow i}\Delta x}{T_{j,i}\cdot w},
$$
which means that for uncertain, bounded $T_{j,i}$, the maximum head drop is bounded as:
\begin{equation}
   \begin{aligned}
     \frac{q_{j\rightarrow i} \cdot \Delta x}{\overline{T_{j,i}} \cdot  w} \leq \Delta h_{\operatorname{j\rightarrow i}} \leq \frac{q_{j\rightarrow i} \cdot \Delta x}{\underline{T_{j,i}} \cdot  w}.
    \end{aligned} 
    \label{apeq:head_drop_bounds}
\end{equation}

If we now consider the head drop over a set of edges $(j,i) \in \operatorname{region}$, the head drop inequality instead becomes:
$$
\begin{aligned}
     \sum_{(j,i) \in \operatorname{region}}\frac{q_{j\rightarrow i} \cdot \Delta x}{\overline{T_{j,i}} \cdot  w} \leq \Delta h_{\operatorname{region}} \leq \sum_{(j,i) \in \operatorname{region}} \frac{q_{j\rightarrow i} \cdot \Delta x}{\underline{T_{j,i}} \cdot  w}.
\end{aligned}
$$

\subsubsection{Center region}

In the center region, we know the head drop is exactly $\Delta h_{\operatorname{center}}=3$~[m]. Reformulating this inequality for $Q$ (we have omitted the subscript because volumetric flow is uniform here) yields:
$$
\begin{aligned}
     \frac{\Delta h_{\operatorname{center}}}{\sum_{(j,i) \in \operatorname{center}}\frac{\Delta x}{\underline{T_{j,i}} \cdot  w}}
     \leq q \leq
     \frac{\Delta h_{\operatorname{center}}}{\sum_{(j,i) \in \operatorname{center}}\frac{\Delta x}{\overline{T_{j,i}} \cdot  w}},
\end{aligned}
$$
which allows us to tighten the flow bounds $\underline{q}$ and $\overline{q}$:
$$
\begin{aligned}
    \underline{q} &= \max(\underline{q}, \frac{\Delta h_{\operatorname{center}}}{\sum_{(j,i) \in \operatorname{center}}\frac{\Delta x}{\underline{T_{j,i}} \cdot  w}}) \\
    \overline{q} &= \min(\overline{q}, \frac{\Delta h_{\operatorname{center}}}{\sum_{(j,i) \in \operatorname{center}}\frac{\Delta x}{\overline{T_{j,i}} \cdot  w}}) \\
\end{aligned}
$$

\subsubsection{Left and right region}

For the left and right regions, the total head drop is not prescribed exactly, but only bounded from above by the admissible head intervals. For the left region, the head at any node must remain below its upper bound, so the cumulative drop between the leftmost part of the domain and the first observation cannot exceed $\Delta h_{\operatorname{left}}^{\max} = 2$~[m]. Likewise, on the right side, the head at any node must remain above its lower bound, so the cumulative drop between the second observation and the rightmost part of the domain cannot exceed $\Delta h_{\operatorname{right}}^{\max} = 4$~[m].

For a given flow rate $q$, Darcy's Law implies that the total head drop over the left region must satisfy
$$
\begin{aligned}
q \sum_{(j,i)\in \operatorname{left}} \frac{\Delta x}{\overline{T_{j,i}}\cdot w}
\leq \Delta h_{\operatorname{left}}
\leq
q \sum_{(j,i)\in \operatorname{left}} \frac{\Delta x}{\underline{T_{j,i}}\cdot w}.
\end{aligned}
$$
At the same time, feasibility of the head bounds requires
$$
0 \leq \Delta h_{\operatorname{left}} \leq \Delta h_{\operatorname{left}}^{\max}.
$$
Hence, a feasible left-region drop exists if and only if these two admissible intervals overlap, which is equivalent to
$$
q \sum_{(j,i)\in \operatorname{left}} \frac{\Delta x}{\overline{T_{j,i}}\cdot w}
\leq \Delta h_{\operatorname{left}}^{\max}.
$$
Reformulating this expression yields the additional upper bound
$$
q \leq
\frac{\Delta h_{\operatorname{left}}^{\max}}
{\sum_{(j,i)\in \operatorname{left}} \frac{\Delta x}{\overline{T_{j,i}}\cdot w}}.
$$

By the same reasoning, in the right region we require
$$
q \sum_{(j,i)\in \operatorname{right}} \frac{\Delta x}{\overline{T_{j,i}}\cdot w}
\leq \Delta h_{\operatorname{right}}^{\max},
$$
which gives
$$
q \leq 
\frac{\Delta h_{\operatorname{right}}^{\max}}
{\sum_{(j,i)\in \operatorname{right}} \frac{\Delta x}{\overline{T_{j,i}}\cdot w}}.
$$

Therefore, the center region tightens both the lower and upper flow bounds, whereas the left and right regions only tighten the upper flow bound. Intuitively, this is because in the center region the total head drop is fixed exactly by the observations, while outside the observation interval we only impose that the cumulative head drop must not become so large that one of the head bounds is violated.

With the tightened flow bounds obtained, we may then sample a concrete volumetric flow rate $q^*$ from a uniform distribution within the initial bounds:
$$
q^*\sim \mathcal{U}(\underline{q},\overline{q}),
$$
and compute the corresponding recharge values $R_1^*$ and $R_{10}^*$ as:
$$
\begin{aligned}
    R_{1}^{*} &= +\frac{q^*}{A} \\
    R_{10}^{*} &= -\frac{q^*}{A} \\
\end{aligned}
$$

\subsection{Sampling $T_{j,i}$ and computing $h_i$}

With a specific flow rate $q^*$ obtained, we make use of the knowledge of the flow direction between source and sink and sample hydraulic head drops $\Delta h_{j \rightarrow i}$ between all the cells in a given $\operatorname{region}$. To this end, we first compute lower and upper bounds for each inter-cell head drop using Equation~\ref{apeq:head_drop_bounds}:
$$
   \begin{aligned}
     \frac{q^* \cdot \Delta x}{\overline{T_{j,i}} \cdot  w} \leq \Delta h_{j\rightarrow i} \leq \frac{q^* \cdot \Delta x}{\underline{T_{j,i}} \cdot  w}.
    \end{aligned} 
$$

For each edge $(j,i) \in \operatorname{region}$, we then sample a head drop $h_{j\rightarrow i}$ uniformly within these bounds, adjusting the lower and upper bounds after each sample to ensure that the regional head drop bounds $h_{\operatorname{region}}$ remain honored. This gives us concrete head drops $h_{j\rightarrow i}^*$ consistent with $q^*$. Combined with the observations, this would allow us to derive the corresponding head values in each cell, but we instead use these head drop values to compute the transmissivities along each edge:
$$
T_{j,i}^* = \frac{q^* \cdot \Delta x}{\Delta h_{j\rightarrow i}^* \cdot w}
$$

Finally, we run a forward FVM simulation using $T_{j,i}^*$, $R_{i}^*$, and the first observation $h_{4}=10$ as a reference head, and obtain the head values $h_i^*$ for each cell. We then repeat this procedure for every sample we want to generate from the exact empirical sampler.

\end{document}